\documentclass[preprint,review,12pt,3p]{elsarticle}
\usepackage{graphicx} 
\usepackage{subfigure}
\usepackage{float} 
\usepackage{amssymb}
\usepackage{color}
\usepackage{amsmath}
\usepackage{mathtools}
\usepackage[thmmarks,amsmath]{ntheorem} 
\usepackage{bm} 
\newcommand{\ud}{\mathrm{d}}
\begin{document}
\begin{frontmatter}

\title{A Continuum Model for Dislocation Climb}

		\author[hkust,whu,shciri]{Chutian Huang}
        \author[whu]{Shuyang Dai\corref{cor1}}
        \cortext[cor1]{Corresponding author}
		\ead{shuyang_dai@whu.edu.cn}
        \author[xut]{Xiaohua Niu}
      \author[hkust]{Tianpeng Jiang}
        \author[whu]{Zhijian Yang}
       \author[ihpc,jhu]{Yejun Gu\corref{cor1}}
		\ead{yejun_gu@ihpc.a-star.edu.sg}
		\author[hkust,shciri]{Yang Xiang\corref{cor1}}
		\ead{maxiang@ust.hk}
		\address[hkust]{Department of Mathematics,
			The Hong Kong University of Science and Technology, Clear Water Bay, Kowloon, Hong Kong}
		\address[whu]{School of Mathematics and Statistics, Wuhan University, Wuhan, Hubei, 430072, China}
         \address[shciri]{HKUST Shenzhen-Hong Kong Collaborative Innovation Research Institute, Futian, Shenzhen,China}
        \address[xut]{School of Mathematics and Statistics, Xiamen University of Technology, Xiamen, 361024, Chin}
         \address[ihpc]{Institute of High Performance Computing (IHPC), Agency for Science, Technology and Research (A*STAR), Singapore 138632, Singapore}
	\address[jhu]{Department of Mechanical Engineering, The Johns Hopkins University, Baltimore, MD 21218, USA}

\begin{abstract}

Dislocation climb plays an important role in understanding plastic deformation of metallic materials at high temperature.
In this paper, we present a continuum formulation for dislocation climb velocity based on densities of dislocations. The obtained continuum formulation is an accurate approximation of the Green's function based discrete dislocation dynamics method ({\it Gu et al. J. Mech. Phys. Solids 83:319-337, 2015}).
The continuum dislocation climb formulation has the advantage of accounting for both the long-range effect of vacancy bulk diffusion and that of the Peach-Koehler climb force, and the two long-range effects are canceled into a short-range effect (integral with fast-decaying kernel) and in some special cases, a completely local effect.
This significantly simplifies the calculation in
the Green's function based discrete dislocation dynamics method, in which a linear system has to be solved over the entire system for the long-range effect of vacancy diffusion and the long-range Peach-Koehler climb force has to be calculated.
 This obtained continuum dislocation climb velocity can be applied in any available continuum dislocation dynamics frameworks.
 We also present numerical validations for this continuum climb velocity and simulation examples for implementation in continuum dislocation dynamics frameworks.


\end{abstract}


 \end{frontmatter}

\section{Introduction}

Dislocations are ubiquitous in metallic materials. Dislocation climb is the motion out of the slip plane assisted by the emission/absorption/transportation  of vacancies or interstitials.
 It has significant contributions to many microscopic/mesoscale plastic mechanism,
 especially at high temperature (e.g., Nabarro-Herring creep and Coble creep)~\cite{poirier1985creep, weertman1957, gao2017, po2022model} or in irradiated materials (e.g., in nuclear fusion and fission reactors)~\cite{dudarev2013,zhu2018,mcelfresh2021,jian2022energetically,Dudarev2022}, due to the enhanced diffusivity and agglomeration of point defects (e.g., self-interstitial atoms, vacancies), respectively.
Mechanisms of dislocation climb
have been extensively investigated through theoretical calculations~\cite{weertman1955theory, nix1971contribution}, atomistic simulations~\cite{kabir2010predicting, fan2013mapping}, and experiments~\cite{li2010situ, chu2022situ}.

Numerous efforts have been undertaken to formulate the coupling between dislocation climb and vacancies/interstitials. For a few simple cases, e.g., an infinitely long straight edge dislocation and a circular prismatic loop, the coupling can be solved analytically and the dislocation climb velocity can be expressed that is proportional to the Peach-Koehler climb force \citep{anderson2017theory}. For general cases having complex dislocation structures, many studies under the framework of discrete dislocation dynamics account for the vacancy-diffusion assisted dislocation climb by adapting the linear relationship between the climb velocity and Peach-Koehler climb force, whose formulation is based upon the equilibrium vacancy distribution for a single, straight edge dislocation~\cite{raabe1998consideration,Ghoniem,xiang2003level,xiang2006dislocation,quek2006,arsenlis2007enabling,mordehai2008introducing,bako2011dislocation,danas2013plane,haghighat2013effect}.
Such expressions drastically lose their accuracy when the dislocations are not sparsely distributed, which prevents from being applied to complex dislocation structures. Some attempts have been made to address this limitation through solving vacancy diffusion equation over the bulk of the materials using finite element methods~\cite{keralavarma2012power,ayas2014climb}.

Alternatively, Gu et al.~\cite{gu2015three} developed a Green's function method to solve for the dislocation climb velocity that is coupled with vacancy diffusion. In this formulation, the dislocation climb velocity is determined from the Peach–Koehler force on dislocations
through vacancy diffusion in a non-local manner through the Green's function of diffusion equilibrium, and the calculations are only limited to the  dislocation nodes rather than over the full three-dimensional domain.
This method considers the long range effects associated with both the Peach-Koehler force and the vacancy bulk diffusion.
This Green's function method provides an accurate and efficient tool in the simulation of dislocation climb. This method has been applied to understand the role of dislocation climb on self-healing~\cite{gu2016relaxation,gu2018self} and point defect sink efficiency~\cite{gu2017point} of low-angle grain boundaries. This nonlocal model has also been extended to include the simultaneous evolution of dislocation loops and cavities in a finite medium~\cite{dudarev2017}.

Although the aforementioned implementations of climb in discrete dislocation dynamics simulations have been utilized to capture the dynamic collective behaviors of dislocation climb, there is still a gap of length/time scales between the practical engineering problems and the discrete dislocation dynamics simulations, which necessitates the development of continuum dislocation dynamics model with explicit and accurate description of dislocation climb. Such continuum dislocation dynamics models provide the basis for the physics-based crystal plasticity theory with a good trade-off between the accuracy (i.e. properly accounting for microstructural information) and efficiency (i.e. capability to be applied to crystals of size reaching tens of microns and beyond) across multiple length and time scales, e.g.~\cite{Acharya2001,Arsenlis2002,groma2003,xiang2009continuum,zhu2010continuum,sandfeld2011continuum,Hochrainer2014,zhuyichao2015continuum,Ngan2014,Niu2018,ngan2022}. Recently, those single-dislocation-based climb velocity expressions have been implemented into some continuum crystal plasticity models to study different physical problems, e.g., thermal and irradiation creep behaviors in steels~\cite{wen2020mechanism}, anisotropy and texture evolution of Mg alloy~\cite{ritzo2022accounting}, creep lifetime predictions in steels~\cite{bieberdorf2021mechanistic}, and a dislocation climb/glide coupled crystal plasticity model with finite element implementation has been proposed~\cite{yuan2018}.
However, to the best of our knowledge, there is no three-dimensional dislocation density based continuum model available in the literature to describe the evolution of dislocation structures by dislocation climb that accurately incorporates the dislocation climb velocity coupled with  vacancy-diffusion process.

%

In this paper, we present  continuum formulations for dislocation climb velocity based on densities of dislocations (Eq.~\eqref{eq:v_cl} for two dimensional problems and Eq.~\eqref{eqn:vcl-3d} for three dimensions). The obtained continuum formulation is an approximation of the Green's function based discrete dislocation dynamics formulation in \cite{gu2015three}. The continuum dislocation climb formulation has the advantage of accounting for both the long-range effect of vacancy bulk diffusion and that of the Peach-Koehler climb force, and the two long-range effects are canceled into a short-range effect (integral with fast-decaying kernel) and in some special cases, a completely local effect.
This significantly simplifies the calculation in
the Green's function based discrete dislocation dynamics method, in which a linear system has to be solved over the entire system for the long-range effect of vacancy diffusion and the long-range Peach-Koehler climb force has to be calculated.
 This continuum climb velocity formulation provides a good approximation for a not very sparse distribution of dislocations, whereas the continuum climb formulation based on mobility law is valid only in the sparse limit of dislocation distributions, as examined with discrete dislocation dynamics model. We also generalize this continuum formulation to include the pipe diffusion-assisted self-climb based on the discrete self-climb dislocation dynamics model~\cite{Niu2017,Niu2019}.  This continuum dislocation climb velocity can be applied in any available continuum dislocation dynamics frameworks, and we present an implementation of this continuum climb formulation in the continuum dislocation dynamics framework using dislocation density potential functions (DDPFs) \cite{zhuyichao2015continuum,Niu2018}.


The rest of this paper is organized as follows. In Sec.~2, we present the formulations within this continuum framework incorporating vacancy bulk diffusion-assisted climb, for both the case of two and three dimensional settings.
In Sec.~3, we present numerical validations of the continuum climb velocity formulation by comparison with the Green's function discrete dislocation climb model, and demonstrate the advantages of obtained continuum climb velocity formulation over the continuum mobility law based climb model.
In Sec.~4, we demonstrate how to incorporation of the obtained continuum dislocation climb velocity formulation in continuum dislocation dynamics models, based on the continuum dislocation dynamics framework using dislocation density potential functions (DDPFs) \cite{zhuyichao2015continuum,Niu2018}.
In the discussion in Sec.~5, we present generalization of the continuum dislocation climb velocity formulation to further  include the self-climb of dislocations in continuum dislocation dynamics models. Conclusions  are drawn in Sec.~6.

\section{Vacancy bulk diffusion-assisted climb in the continuum framework}\label{sec:2}
In this section, we establish a continuum dislocation climb velocity formulation based on the discrete dislocation plasticity coupled with the vacancy bulk diffusion. This continuum dislocation climb velocity can be applied in any available continuum dislocation dynamics frameworks.

We first briefly review two climb models under the framework of discrete dislocation dynamics.
The mobility law climb velocity formulation in a discrete dislocation dynamics model \cite{anderson2017theory} is based on a vacancy-assisted climb of a single edge dislocation
\begin{equation}
v_{\rm cl}=\frac{2\pi c_0D_v \Omega}{b^2k_BT\ln(r_\infty/r_d)}f_{\rm cl}.\label{eqn:mobility-law}
\end{equation}
It has been used in many discrete and dislocation density based simulations as reviewed in the introduction section, where in a dislocation density based model, the climb force $f_{\rm cl}$ in this mobility law form is calculated from dislocation densities. Note that climb velocity $v_{\rm cl}$ is in the direction of $\pmb \xi\times \mathbf b/b$, where $\pmb \xi$ is the dislocation line direction,  and $\mathbf b$ is the Burgers vector with length $b$, i.e.
\begin{equation}
\mathbf v_{\rm cl}=v_{\rm cl}(\pmb \xi\times \mathbf b/b).
\end{equation}

The Greens's function method for dislocation climb under the framework of discrete dislocation dynamics~\cite{gu2015three} takes into account the long-range effect of vacancy bulk diffusion, which is neglected in the local mobility law climb formulation.
Below is a briefly review of this Greens's function method in discrete dislocation dynamics, based on which our continuum dislocation climb velocity formulation will be derived. Assuming vacancy diffusion equilibrium, the vacancy concentration $c$ satisfies
\begin{align}
&D_v \nabla^2 c=b_e v_{\rm cl}\delta(\Gamma),
\label{eq:governing-d}
\end{align}
where $v_{\rm cl}$ is the dislocation climb velocity, $b_e$ is the edge component of the Burgers vector, $D_v$ is the vacancy bulk diffusion, and   $\Gamma$ refers to all of the dislocation lines in the entire system. This diffusion equilibrium equation is subject to the constant vacancy concentration boundary condition at far fields: $c(r=r_\infty) = c_{\infty}$. The dislocation climb velocity $v_{\rm cl}$ in the Greens's function method \cite{gu2015three} is obtained by solving the following system of integral equations along the dislocations:
\begin{equation}
\sum_{\gamma_j\subset \Gamma}\frac{1}{D_v}\int_{\gamma_j}G(x_1-x,y_1-y,z_1-z)b_e(x_1,y_1,z_1)v_{\rm cl}(x_1,y_1,z_1)
\ud l+c_\infty=c_0 \exp\left(-\frac{f_{\rm cl}\Omega}{b_e k_{\rm B} T}\right)\label{eqn:discrete000}
\end{equation}
for any point $(x,y,z)$ on a dislocation $\gamma_i\subset \Gamma$. Here $G(x,y,z)=-\frac{1}{4\pi}\frac{1}{\sqrt{x^2+y^2+z^2}}$ is the Green's function of the Laplace equation in three dimensions, $(x_1,y_1,z_1)$ in the integral varies along the dislocation $\gamma_j$, $\ud l$ is the line element of the integral,
 $f_{\rm cl}$ is the climb force, $c_0$ is the vacancy equilibrium concentration without the climb force, $\Omega$ is the atomic volume, $k_{\rm B}$ is the Boltzmann constant, and $T$ is the temperature.

 Note that since Eq.~\eqref{eqn:discrete000} comes from the condition on the surface of the dislocation tube with radius $r = r_d$, the Green's function $G$ in it can be approximated by $G(x_1-x,y_1-y,z_1-z)\approx -\frac{1}{4\pi}\frac{1}{\sqrt{(x_1-x)^2+(y_1-y)^2+(z_1-z)^2+r_d}}$ when we want to avoid singular integrals in Eq.~\eqref{eqn:discrete000}.

When all the dislocations are straight edge dislocations, this Green's function method is reduced to a two-dimensional formulation in a cross-section plane:
\begin{equation}
\sum_{j=1}^N\frac{b }{D_v}G(x^{(j)}-x^{(i)}, y^{(j)}-y^{(i)})v_{\rm cl}^{(i)}+c_\infty=c_0 \exp\left(-\frac{f_{\rm cl}\Omega}{b_e k_{\rm B} T}\right), \ \ i=1,2,\cdots,N.
\label{eqn:discrete0002d}
\end{equation}
Here $v_{\rm cl}^{(i)}$ is the climb velocity of the $i$-th dislocation, which is located at $(x^{(i)}, y^{(i)})$,
$G(x,y)$ is the Green's function of the Laplace equation in two dimensions, $G(x,y)\approx\frac{1}{2\pi}\ln\frac{\sqrt{x^2+y^2}}{r_\infty}$ for $(x,y)\neq(0,0)$ and $G(0,0)\approx\frac{1}{2\pi}\ln\frac{r_d}{r_\infty}$, where $r_\infty$ is an outer cutoff radius and $r_\infty\gg r_d$.
When there is only a single dislocation in the system, this formulation reduces to the mobility law in Eq.~\eqref{eqn:mobility-law}; for multiple dislocations, this formulation incorporates the long-range effect of vacancy bulk diffusion, and the mobility law is not able to provide good approximation in this case due to neglect of this long-range effect \cite{gu2015three}.

\subsection{Two-dimensional (2D) continuum formulation}\label{sec:2D}
The governing vacancy bulk diffusion equation can be established in the continuum framework with respect to the dislocation climb velocity ($v_{\rm cl}$, under the assumption of quickly equilibrated vacancy diffusion in bulk (i.e., $\frac{\partial c}{\partial t}=0$), similar to the discrete vacancy diffusion-assisted dislocation climb model,
\begin{align}
&D_v \nabla^2 c=b v_{\rm cl}(x,y)\rho(x,y),
\label{eq:governing}
\end{align}
subject to the constant vacancy concentration boundary condition at far fields,
\begin{align}
    c(r=r_\infty) = c_{\infty},
\label{eq:bc_far}
\end{align}
where $\mathbf b=(b,0)$ is the Burgers vector,  $\rho$ is the dislocation density,  and $v_{\rm cl}$ is the continuum climb velocity.

Now we consider the solution to the boundary value problem, Eqs.~\eqref{eq:governing} and~\eqref{eq:bc_far},. Using Green's function of the 2D Laplace equation, $G(x,y)=\frac{1}{2\pi}\ln\frac{\sqrt{x^2+y^2}}{r_\infty}$, the vacancy concentration as described by Eqs.~\eqref{eq:governing} and \eqref{eq:bc_far} can be expressed as the convolution of the Green's function and average climb:
\begin{align}
c(x,y)=\frac{b}{D_v} G*(v_{\rm cl}\rho)+c_\infty.
\label{eq:G_2d}
\end{align}
where $*$ denotes the convolution operation, i.e., $f*g(x,y)=\int_{\mathbb{R}^2}f(x-x_1,y-y_1)g(x_1,y_1)\ud x_1 \ud y_1$.

On the other hand, Eq.~\eqref{eq:G_2d} satisfies chemical equilibrium condition everywhere in the domain
\begin{align}
    c(x,y)=c_0 \exp\left(-\frac{f_{\rm cl}\Omega}{b k_{\rm B} T}\right),
    \label{eq:bc_core}
\end{align}
Usually, with the assumption that $f_{\rm cl}\ll bk_{\rm B}T/\Omega$, Eq.~\eqref{eq:bc_core} can be linearized as
\begin{align}
    c(x,y)\approx c_0\left[1-\frac{\Omega}{b k_{\rm B} T}\cdot f_{\rm cl}(x,y)\right].
    \label{eq:bc_core2}
\end{align}

In the continuum model, the climb force at the point $(x,y)$,  is
\begin{align}
f_{\rm cl}(x,y)=\frac{\mu b^2}{2\pi(1-\nu)}\int_{\mathbb{R}^2}\frac{(y_1-y)[3(x_1-x)^2+(y_1-y)^2]}{[(x_1-x)^2+(y_1-y)^2]^2}\cdot \rho(x_1,y_1)\ud x_1\ud y_1+f_{\rm cl}^0(x,y),
\label{eq:f_cl}
\end{align}
where the first term is the climb force due to the stress generated by all the dislocations, $\mu$ is the shear modulus, $\nu$ is the Poisson's ratio, and $f_{\rm cl}^0(x,y)$ is the climb force due to other stress fields, e.g. the applied stress. 
This continuum formulation for the climb force is  based on the discrete dislocation model~\citep{anderson2017theory}:
\begin{align}
f_{\rm cl}^{(i_0,j_0)}=\frac{\mu b^2}{2\pi(1-\nu)}\sum_{i\neq i_0,\ j\neq j_0}\frac{(y^{(i,j)}-y^{(i_0,j_0)})[3(x^{(i,j)}-x^{(i_0,j_0)})^2+(y^{(i,j)}-y^{(i_0,j_0)})^2]}
{[(x^{(i,j)}-x^{(i_0,j_0)})^2+(y^{(i,j)}-y^{(i_0,j_0)})^2]^2}+f_{\rm cl}^{0,(i_0,j_0)},
\label{eq:f_cl-d}
\end{align}
where $f_{\rm cl}^{(i,j)}$ is the climb force on the $(i,j)$-th dislocation, $(x^{(i,j)},y^{(i,j)})$ is the location of the $(i,j)$-th dislocation, and $f_{\rm cl}^{0,(i,j)}$ is the climb force  on the $(i,j)$-th dislocation due to the applied stress.

The continuum climb force of Eq.~\eqref{eq:f_cl} can be expressed in the convolution form,
\begin{align}
f_{\rm cl}(x,y)&=\frac{\mu b^2}{2\pi(1-\nu)}H*\rho(x,y)+f_{\rm cl}^0(x,y),
\label{eq:f_cl_conv}
\end{align}
where $H(x,y)=\frac{y(3x^2+y^2)}{(x^2+y^2)^2}$.

Substituting Eq.~\eqref{eq:f_cl_conv} into the combined Eqs.~\eqref{eq:G_2d} and~\eqref{eq:bc_core2}, the relationship between the average climb velocity and the climb force is described as
\begin{align}
G*(v_{\rm cl}\rho)= -\frac{c_0\mu D_v\Omega}{2\pi(1-\nu) k_{\rm B} T} H*\rho-\frac{c_0D_v\Omega}{b^2 k_{\rm B} T}f_{\rm cl}^0-\frac{D_v(c_\infty-c_0)}{b}.
\label{eq:G_2d2}
\end{align}


According to the convolution theorem, the Fourier transforms  of Eqs.~\eqref{eq:f_cl_conv} and \eqref{eq:G_2d2}  read
\begin{flalign}
\widehat{f_{\rm cl}}=&\frac{2\pi\mu b^2}{1-\nu}\hat{H}\cdot\hat{\rho}+\widehat{f_{\rm cl}^0}\label{eqn:fft-fcl10}\\
\hat{G}\cdot\widehat{(v_{\rm cl}\rho)}=&-\frac{c_0D_v\mu\Omega}{2\pi (1-\nu)k_{\rm B} T}\hat{H}\cdot\hat{\rho}-\frac{c_0D_v\Omega}{4\pi^2b^2 k_{\rm B} T}\widehat{f_{\rm cl}^0}-\widehat{\frac{D_v(c_\infty-c_0)}{4\pi^2 b}}.
\label{eq:G_2d3}
\end{flalign}
Recall that the Fourier transform of the function $f$, $\mathcal{F}(f)$, and corresponding inverse Fourier transform, $\mathcal{F}^{-1}(\mathcal{F}(f))$, are defined as
\begin{align}
\hat{f}(k_1,k_2)=&\mathcal{F}(f)(k_1,k_2)=\frac{1}{4\pi^2}\int_{\mathbb{R}^2} f(x,y)e^{-i(k_1x+k_2y)}\ud x\ud y,\\
f(x,y)=&\mathcal{F}^{-1}(\hat{f})(x,y)=\int_{\mathbb{R}^2}\hat{f}(k_1,k_2)e^{i(k_1x+k_2y)}\ud k_1 \ud k_2.
\end{align}

Following Fourier transforms $\hat{G}(k_1,k_2)=-\frac{1}{4\pi^2(k_1^2+k_2^2)}$ and $\hat{H}(k_1,k_2)=-\frac{ik_2^3}{\pi (k_1^2+k_2^2)^2}$
it can be  calculated from Eqs.~\eqref{eqn:fft-fcl10} and  \eqref{eq:G_2d3} that
\begin{flalign}
\widehat{f_{\rm cl}}=&-\frac{2\mu b^2}{1-\nu}\frac{ik_2^3}{ (k_1^2+k_2^2)^2}\hat{\rho}+\widehat{f_{\rm cl}^0},\label{eqn:fft-fcl20}
\end{flalign}
and
\begin{flalign}
\widehat{v_{\rm cl}\rho}=&\frac{2c_0D_v\mu\Omega}{(1-\nu)k_{\rm B} T}\cdot\frac{ik_2^3}{k_1^2+k_2^2} \hat{\rho}+\frac{c_0D_v\Omega}{b^2 k_{\rm B} T}(k_1^2+k_2^2)\widehat{f_{\rm cl}^0}, \ \ {\rm for}\ k_1^2+k_2^2\neq 0, \label{eq:sol}\\
\overline{v_{\rm cl}\rho}=&\frac{4 c_0D_v\Omega}{b^2 k_{\rm B} T r_\infty^2}\left(\overline{f_{\rm cl}^0}+\frac{b k_{\rm B} T}{c_0\Omega}
(c_\infty-c_0)\right). \label{eq:sol-0}
\end{flalign}
Here  Eq.~\eqref{eq:sol-0} is for $k_1=k_2= 0$, and $\overline{g}=\frac{1}{\pi r_\infty^2}\int_{r\leq r_\infty} g(x,y)\ud x\ud y$ is the average value of $g(x,y)$.


We introduce a function $J(x,y)=\frac{2\pi y}{x^2+y^2}$, and we have  $\hat{J}=-\frac{ik_2}{k_1^2+k_2^2}$.  Thus the first term in Eq.~\eqref{eq:sol} becomes
$\frac{2c_0D_v\mu\Omega}{(1-\nu)k_{\rm B} T}\widehat{\frac{\partial^2 J}{\partial y^2}}\hat{\rho}$.
Performing the inverse Fourier transform in Eqs.~\eqref{eq:sol} and \eqref{eq:sol-0}, and approximating $\rho$ as a constant in Eq.~\eqref{eq:sol-0}, we have
the expression of the continuum climb velocity
\begin{flalign}
    v_{\rm cl} (x,y)
    =&\frac{c_0D_v\mu\Omega}{\pi(1-\nu)k_{\rm B} T}\frac{1}{\rho}\frac{\partial^2}{\partial y^2}\left(\int_{\mathbb{R}^2}\frac{y_1-y}{(x_1-x)^2+(y_1-y)^2} \rho(x_1,y_1)\ud x_1\ud y_1\right)\vspace{1ex}\nonumber\\
    &+\frac{c_0D_v\Omega}{b^2 k_{\rm B} T}\frac{1}{\rho}\nabla^2 f_{\rm cl}^0+\frac{4 c_0D_v\Omega}{b^2 k_{\rm B} T r_\infty^2}\frac{1}{\rho}\left(\overline{f_{\rm cl}^0}+\frac{b k_{\rm B} T}{c_0\Omega}
(c_\infty-c_0)\right).
    \label{eq:v_cl}
\end{flalign}


The continuum climb velocity formulation, Eq.~\eqref{eq:v_cl}, as an continuum approximation of the Green's function based discrete dislocation dynamics formulation in \cite{gu2015three}, has the advantage of accounting for both the long-range nature of vacancy diffusion and that of the Peach-Koehler climb force, and the two long-range effects are canceled into a short-range effect (the first, integral term in \eqref{eq:v_cl}, whose integral kernel is fast-decaying as discussed below).
This significantly simplifies the calculation in
the Green's function based discrete dislocation dynamics method, in which a linear system has to be solved over the entire system for the long-range effect of vacancy diffusion and the long-range Peach-Koehler climb force has to be calculated.

 Note that the integral in the first term in Eq.~\eqref{eq:v_cl} can be written as $$\frac{\partial^2}{\partial y^2}\left(\int_{\mathbb{R}^2}\frac{y_1-y}{(x_1-x)^2+(y_1-y)^2} \rho(x_1,y_1)\ud x_1\ud y_1\right)
=\frac{\partial^2 J}{\partial y^2}*\rho,$$ where $J(x,y)=\frac{2\pi y}{x^2+y^2}$. Since $\frac{\partial^2 J}{\partial y^2}$ decays fast as $x,y\rightarrow +\infty$, this integral can be calculated by cutting off to one over a finite neighborhood of the point $(x,y)$.

This continuum climb velocity formulation provides a good approximation for a not very sparse distribution of dislocations, whereas the continuum climb formulation based on mobility law in Eq.~\eqref{eqn:mobility-law} is valid only in the sparse limit of dislocation distributions; see the examples in Sec.~\ref{sec:validation}.

Finally, we consider the continuum climb velocity formulation \eqref{eq:v_cl} for a special case.

\underline{\bf 2D dislocation distribution uniform in one direction}.

\begin{figure}[ht]
\centering
  \includegraphics[width=0.5\textwidth]{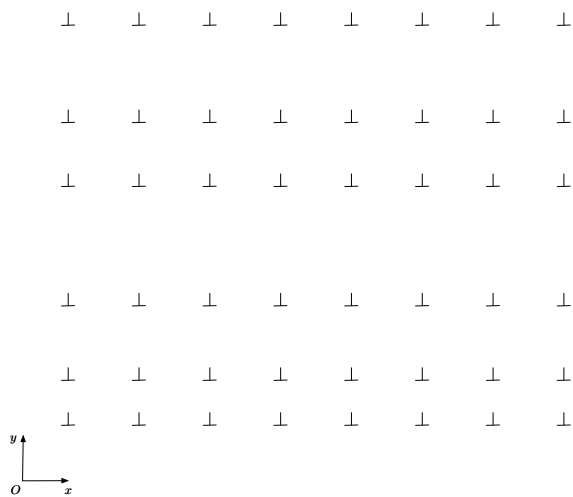}
  \caption{2D dislocation distribution uniform in one direction.}
  \label{fig:x_uni}
\end{figure}

Considering a 2D case where the dislocation density is uniform along the $x$-direction and varies along the $y$-direction, as shown in Fig.~\ref{fig:x_uni}.
It is reduced to a one-dimensional (1D) problem with the variable $y$. 

In the case with the assumption that $c_\infty=c_0$ and $f_{\rm cl}^0=0$, the continuum climb velocity in Eq.~\eqref{eq:v_cl} gives
\begin{align}
v(y)&=\frac{c_0D_v\mu\Omega}{\pi(1-\nu)k_{\rm B} T}\frac{1}{\rho}\frac{\partial^2}{\partial y^2}\int_{\mathbb{R}^2}\frac{y_1-y}{(x_1-x)^2+(y_1-y)^2}\rho(y_1)\ud x_1\ud y_1\nonumber\\
&=\frac{ c_0D_v\mu\Omega}{(1-\nu)k_{\rm B} T}\frac{1}{\rho}\frac{\ud^2}{\ud y^2}\int_{-\infty}^{+\infty}{\rm sgn}(y_1-y)\rho(y_1)\ud y_1\nonumber\\
&=\frac{ c_0D_v\mu\Omega}{(1-\nu)k_{\rm B} T}\frac{1}{\rho}\frac{\ud^2}{\ud y^2}\left(-\int_{-\infty}^{y}\rho(y_1)\ud y_1+\int_{y}^{+\infty}\rho(y_1)\ud y_1\right)\nonumber\\
&=-\frac{ c_0D_v\mu\Omega}{(1-\nu)k_{\rm B} T}\frac{1}{\rho}\frac{\ud}{\ud y}(2\rho(y))\nonumber\\
&=-\frac{2 c_0D_v\mu\Omega}{(1-\nu)k_{\rm B} T}\cdot\frac{\rho'(y)}{\rho(y)}.
\label{eq:1D}
\end{align}

Note that in this special case, the climb velocity formulation is local, i.e., there is no integral in it. The physical meaning is that the two long-range effects cancel out in this special case.

On the other hand, the mobility law climb velocity in Eq.~\eqref{eqn:mobility-law} for this case is
\begin{equation}
v_{\rm cl}=\frac{2\pi c_0D_v \Omega}{b^2k_BT\ln(r_\infty/r_d)}f_{\rm cl}=\frac{2\pi c_0D_v\mu \Omega}{(1-\nu)k_BT\ln(r_\infty/r_d)}\int_{-\infty}^{+\infty}{\rm sgn}(y_1-y)\rho(y_1)\ud y_1.\label{eqn:mobility-law000}
\end{equation}
This is quite different from our continuum climb velocity in Eq.~\eqref{eq:1D}, and it has a long-range effect inherited from the climb force. We will show in the second numerical example in the next section that our continuum climb formulation in Eq.~\eqref{eq:1D} provides an accurate approximation to the discrete model, whereas the mobility law formulation in Eq.~\eqref{eqn:mobility-law000} fails to give correct approximation to the discrete model.

%
%


\subsection{Three-dimensional (3D) continuum formulation}\label{subsection:3D}
Now we consider a more general 3D configuration. We focus on dislocations in the same slip systems, which share the same Burgers vector $\bm{b}=(b,0,0)$ and the same slip plane normal $(0,0,1)$. The climb direction is accordingly the same as the slip plane normal. For a dislocation in this slip system with line direction $\pmb \xi=(\xi_1,\xi_2,\xi_3)$, the edge component of the Burgers vector is $b_e=|b\xi_2|$. The continuum climb velocity formulation will be derived based on this slip system with the dislocation lines being subject to small perturbations  out of the slip planes.


The governing diffusion equation for the 3D case is
\begin{align}
&D_v  \nabla^2 c=b_e v_{\rm cl}(x,y,z)\rho(x,y,z),
\label{eq:governing3D}
\end{align}
subject to far field boundary condition,
\begin{align}
    c(r=r_\infty) = c_{\infty}.
\label{eq:bc_far3d}
\end{align}
Again, the solution to this boundary value problem satisfies the chemical potential equilibrium condition, under the assumption that $f_{\rm cl}\ll bk_{\rm B}T/\Omega$:
\begin{align}
    c(x,y,z)=c_0 \exp\left(-\frac{f_{\rm cl}\Omega}{b_e k_{\rm B} T}\right)\approx c_0\left[1-\frac{\Omega}{b_e k_{\rm B} T}\cdot f_{\rm cl}(x,y,z)\right].
    \label{eq:bc_core3d}
\end{align}
Here the leading order climb force is~\cite{gu2016relaxation}
\begin{align}
f_{\rm cl}=\sigma_{11}b_e+f_{\rm cl}^0.
\end{align}
where $f_{\rm cl}^0$ is the climb force due to the applied loading, and  $\sigma_{11}$ is the stress component having a simple FFT formulation~\cite{xiang2003level,zhu2014continuum}
\begin{align}
    \hat{\sigma}_{11}=\frac{2\mu b}{1-\nu}\frac{i(k_2^2+k_3^2)}{(k_1^2+k_2^2+k_3^2)^2}[k_3\widehat{(\rho\xi_2)}-k_2\widehat{(\rho\xi_3)}].
\end{align}
Combining Eqs~\eqref{eq:governing3D},~\eqref{eq:bc_far3d} and~\eqref{eq:bc_core3d}, using the leading order approximation $b_e\approx b$,   we have
\begin{align}
G*(v_{\rm cl}\rho|\xi_2|)=-\frac{c_0\Omega D_v}{bk_{\rm B}T}\sigma_{11}-\frac{c_0D_v\Omega}{b^2 k_{\rm B} T}f_{\rm cl}^0-\frac{D_v(c_\infty-c_0)}{b},
\label{eq:v_cl3d}
\end{align}
where $G(x,y,z)=-\frac{1}{4\pi\sqrt{x^2+y^2+z^2}}$ is the Green's function of the Laplace equation in 3D.

Performing Fourier transform on both sides of Eq.~\eqref{eq:v_cl3d} leads to
\begin{align}
    8\pi^3\widehat{G} \cdot \widehat{(v_{\rm cl}\rho|\xi_2|)}&=-\frac{c_0 D_v \Omega}{b k_{\rm B} T}\widehat{\sigma}_{11}-\frac{c_0D_v\Omega}{4\pi^2b^2 k_{\rm B} T}\widehat{f_{\rm cl}^0}-\widehat{\frac{D_v(c_\infty-c_0)}{4\pi^2 b}}.
    \label{eq:v_clconv1}
\end{align}
Given that $\widehat{G}=-\frac{1}{8\pi^3(k_1^2+k_2^2+k_3^2)}$, it can be calculated from Eq.~\eqref{eq:v_clconv1} that
\begin{flalign}
    \widehat{v_{\rm cl}\rho|\xi_2|}=&\frac{2\mu c_0 \Omega D_v}{(1-\nu)k_{\rm B} T}\cdot\frac{i(k_2^2+k_3^2)}{k_1^2+k_2^2+k_3^2}\left[k_3\widehat{(\rho\xi_2)}-k_2\widehat{(\rho\xi_3)}\right]+\frac{c_0D_v\Omega}{b^2 k_{\rm B} T}(k_1^2+k_2^2+k_3^2)\widehat{f_{\rm cl}^0}, \label{eq:v_clconv2}\\
    &\ \ {\rm for}\ k_1^2+k_2^2+k_3^2\neq 0, \nonumber\\
    \overline{v_{\rm cl}\rho|\xi_2|}=&\frac{2 c_0D_v\Omega}{b^2 k_{\rm B} T r_\infty^2}\left(\overline{f_{\rm cl}^0}+\frac{b k_{\rm B} T}{c_0\Omega}
(c_\infty-c_0)\right).
    \label{eq:v_clconv2-000}
\end{flalign}
Here  Eq.~\eqref{eq:v_clconv2-000} is for $k_1=k_2=k_3= 0$, and $\overline{g}$ is the average value of $g(x,y,z)$ over $r\leq r_\infty$.

Performing inverse Fourier transform in Eqs.~\eqref{eq:v_clconv2} and \eqref{eq:v_clconv2-000},   the 3D climb velocity is expressed as
\begin{flalign}
    v_{\rm cl}\rho|\xi_2|= &\frac{\mu c_0 \Omega D_v}{2\pi(1-\nu)k_{\rm B} T}
    \left(\frac{\partial^2}{\partial y^2}+\frac{\partial^2}{\partial z^2}\right)
    \left[\frac{z}{(x^2+y^2+z^2)^{\frac{3}{2}}}*(\rho\xi_2)-\frac{y}{(x^2+y^2+z^2)^{\frac{3}{2}}}*(\rho\xi_3)\right]\nonumber\\
    &+\frac{c_0D_v\Omega}{b^2 k_{\rm B} T}\nabla^2 f_{\rm cl}^0+\frac{2 c_0D_v\Omega}{b^2 k_{\rm B} T r_\infty^2}\left(\overline{f_{\rm cl}^0}+\frac{b k_{\rm B} T}{c_0\Omega}
(c_\infty-c_0)\right).
\end{flalign}
Thus, the continuum climb velocity in 3D is
\begin{align}
    v_{\rm cl}(x,y,z) =& \frac{\mu c_0 \Omega D_v}{2\pi(1-\nu)k_{\rm B} T}\frac{b}{ b_e \rho}
    \left(\frac{\partial^2}{\partial y^2}+\frac{\partial^2}{\partial z^2}\right)\nonumber\\
    &\int_{\mathbb{R}^3}\left[\frac{z_1-z}{((x_1-x)^2+(y_1-y)^2+(z_1-z)^2)^{\frac{3}{2}}}\rho(x_1,y_1,z_1)\xi_2(x_1,y_1,z_1)\right.\nonumber\\
    &\left.-\frac{y_1-y}{((x_1-x)^2+(y_1-y)^2+(z_1-z)^2)^{\frac{3}{2}}}\rho(x_1,y_1,z_1)\xi_3(x_1,y_1,z_1)\right]\ud x_1\ud y_1\ud z_1\nonumber\\
    &+\frac{c_0D_v\Omega}{bb_e k_{\rm B} T}\frac{1}{\rho}\nabla^2 f_{\rm cl}^0+\frac{2 c_0D_v\Omega}{bb_e k_{\rm B} T r_\infty^2}\frac{1}{ \rho}\left(\overline{f_{\rm cl}^0}+\frac{b k_{\rm B} T}{c_0\Omega}
(c_\infty-c_0)\right).
    \label{eqn:vcl-3d}
\end{align}

This continuum climb velocity formulation, as the 2D model, has the advantage of accounting for both the long-range effects of vacancy diffusion and the Peach-Koehler climb force, and the two long-range effects are canceled into a short-range effect (the first integral term in \eqref{eqn:vcl-3d}, whose integral kernel is fast-decaying as discussed below).
This significantly simplifies the calculation in
the Green's function based discrete dislocation dynamics method~\cite{gu2015three}, in which a linear system has to be solved over the entire system for the long-range effect of vacancy diffusion and the long-range Peach-Koehler climb force has to be calculated.
 Recall that this continuum model is derived from small perturbations in parallel straight dislocation arrays. It
 provides a good approximation in dislocation density based continuum models, for a not very sparse distribution of dislocations in which the dislocations locally can be considered as almost straight ones.

 Note that the integral in the first term in Eq.~\eqref{eqn:vcl-3d}, without the constant coefficient, can be written as
$$-(\frac{\partial^3}{\partial y^2\partial z}+\frac{\partial^3}{\partial z^3})\frac{1}{r}*(\rho \xi_2)
+(\frac{\partial^3}{\partial y^3}+\frac{\partial^3}{\partial y\partial z^2})\frac{1}{r}*(\rho \xi_3).$$
 Since the third partial derivatives of $1/r$ decay fast as $r\rightarrow +\infty$, this integral can be calculated by cutting off to one over a finite domain, as in the 2D case.

The 3D continuum climb formulation in Eq.~\eqref{eqn:vcl-3d} reduces to the 2D formulation Eq.~\eqref{eq:v_cl}, when $\pmb \xi=(0,1,0)$ and $b_e=b$.
This continuum climb velocity formulation applies to dislocations of a single slip system. It can be used in dislocation density based continuum dislocation dynamics models for the averaged climb behavior, i.e. climb of the geometrically necessary dislocations.

Below we consider the continuum climb velocity formulation Eq.~\eqref{eqn:vcl-3d} for a special case.

\underline{\bf 3D dislocation distribution uniform in one direction}.

Considering a 3D case where the dislocation density is uniform along the $x$-direction, i.e., in the direction of Burgers vector.
It is reduced to a 2D problem with Burgers vector normal to the plane that contains the dislocations (prismatic dislocation loops/lines).

In this case, when $c_\infty=c_0$ and $f_{\rm cl}^0=0$, the continuum climb velocity in Eq.~\eqref{eqn:vcl-3d} gives
\begin{align}
   v_{\rm cl}(y,z) =& \frac{\mu c_0 \Omega D_v}{2\pi(1-\nu)k_{\rm B} T}\frac{1}{\rho}
    \left(\frac{\partial^2}{\partial y^2}+\frac{\partial^2}{\partial z^2}\right)\nonumber\\
    &\int_{\mathbb{R}^3}\left[\frac{z_1-z}{((x_1-x)^2+(y_1-y)^2+(z_1-z)^2)^{\frac{3}{2}}}\rho(y_1,z_1)\xi_2(y_1,z_1)\right.\nonumber\\
    &\left.-\frac{y_1-y}{((x_1-x)^2+(y_1-y)^2+(z_1-z)^2)^{\frac{3}{2}}}\rho(y_1,z_1)\xi_3(y_1,z_1)\right]\ud x_1\ud y_1\ud z_1\nonumber\\
=& \frac{\mu c_0 \Omega D_v}{\pi(1-\nu)k_{\rm B} T}\frac{1}{\rho}
    \left(\frac{\partial^2}{\partial y^2}+\frac{\partial^2}{\partial z^2}\right)\nonumber\\
    &\int_{\mathbb{R}^2}\left[\frac{z_1-z}{(y_1-y)^2+(z_1-z)^2}\rho(y_1,z_1)\xi_2(y_1,z_1)\right.\nonumber\\
    &\left.-\frac{y_1-y}{(y_1-y)^2+(z_1-z)^2}\rho(y_1,z_1)\xi_3(y_1,z_1)\right]\ud y_1\ud z_1\nonumber\\
=& \frac{2\mu c_0 \Omega D_v}{(1-\nu)k_{\rm B} T}\frac{1}{ \rho}
    \left(\frac{\partial^2}{\partial y^2}+\frac{\partial^2}{\partial z^2}\right)\nonumber\\
    &\int_{\mathbb{R}^2}\left[\frac{\partial}{\partial z}\left(\frac{1}{2\pi}\ln\sqrt{(y_1-y)^2+(z_1-z)^2}\right)\rho(y_1,z_1)\xi_2(y_1,z_1)\right.\nonumber\\
    &\left.-\frac{\partial}{\partial y}\left(\frac{1}{2\pi}\ln\sqrt{(y_1-y)^2+(z_1-z)^2}\right)\rho(y_1,z_1)\xi_3(y_1,z_1)\right]\ud y_1\ud z_1\nonumber\\
=& \frac{2\mu c_0 \Omega D_v}{(1-\nu)k_{\rm B} T}\frac{1}{ \rho}
    \left[\frac{\partial}{\partial z}\big(\rho\xi_2\big)-\frac{\partial}{\partial y}\big(\rho\xi_3\big)\right].
    \label{eq:3D111}
\end{align}

Note that in this special case, the climb velocity formulation is local, i.e., there is no integral in it. The physical meaning is that the two long-range effects cancel out in this special case.
Such cancellation of the two long-range effects has been observed in the analysis of relaxation of perturbed low-angle grain boundary by vacancy assisted climb of the constituent dislocations \cite{gu2016relaxation}.

\section{Numerical Validations}\label{sec:validation}

\subsection{Uniform 2D array of straight edge dislocations}
We first consider a uniform 2D array of straight edge dislocations, where dislocations are located at $(x,y)=(iB, jD)$ with $B$ and $D$ being the inter-dislocation distances in the $x$ and $y$ directions, respectively, and $i$, $j$ are integers. We compare the results obtained using our continuum climb velocity formulation in Eq.~\eqref{eq:v_cl}, the discrete Green's function method \cite{gu2015three} given in Eq.~\eqref{eqn:discrete0002d}, and the continuum mobility law formulation in Eq.~\eqref{eqn:mobility-law}.

%
We consider the simulation domain $[0,L_1]\times[0,L_2]$ with periodic boundary conditions. We set $L_1=L_2=3000b$. We set the outer cutoff $r_\infty=\frac{1}{2}\sqrt{L_1^2+L_2^2}$, and the dislocation core cutoff $r_d=2b$. The dislocation density is $\rho=\frac{1}{BD}$. We set $B=D$ and vary their values from $3000b$ down to $100b$. We set $c_\infty=c_0$ for the vacancy diffusion. A constant climb force $f_{\rm cl}^0=0.01 \mu b^2/2\pi(1-\nu)$ is applied. In this case, in the continuum climb velocity in  Eq.~\eqref{eq:v_cl}, only the $\overline{f_{\rm cl}^0}=f_{\rm cl}^0$ term is nonzero.

\begin{figure}[ht]
\centering
  \includegraphics[width=0.7\textwidth]{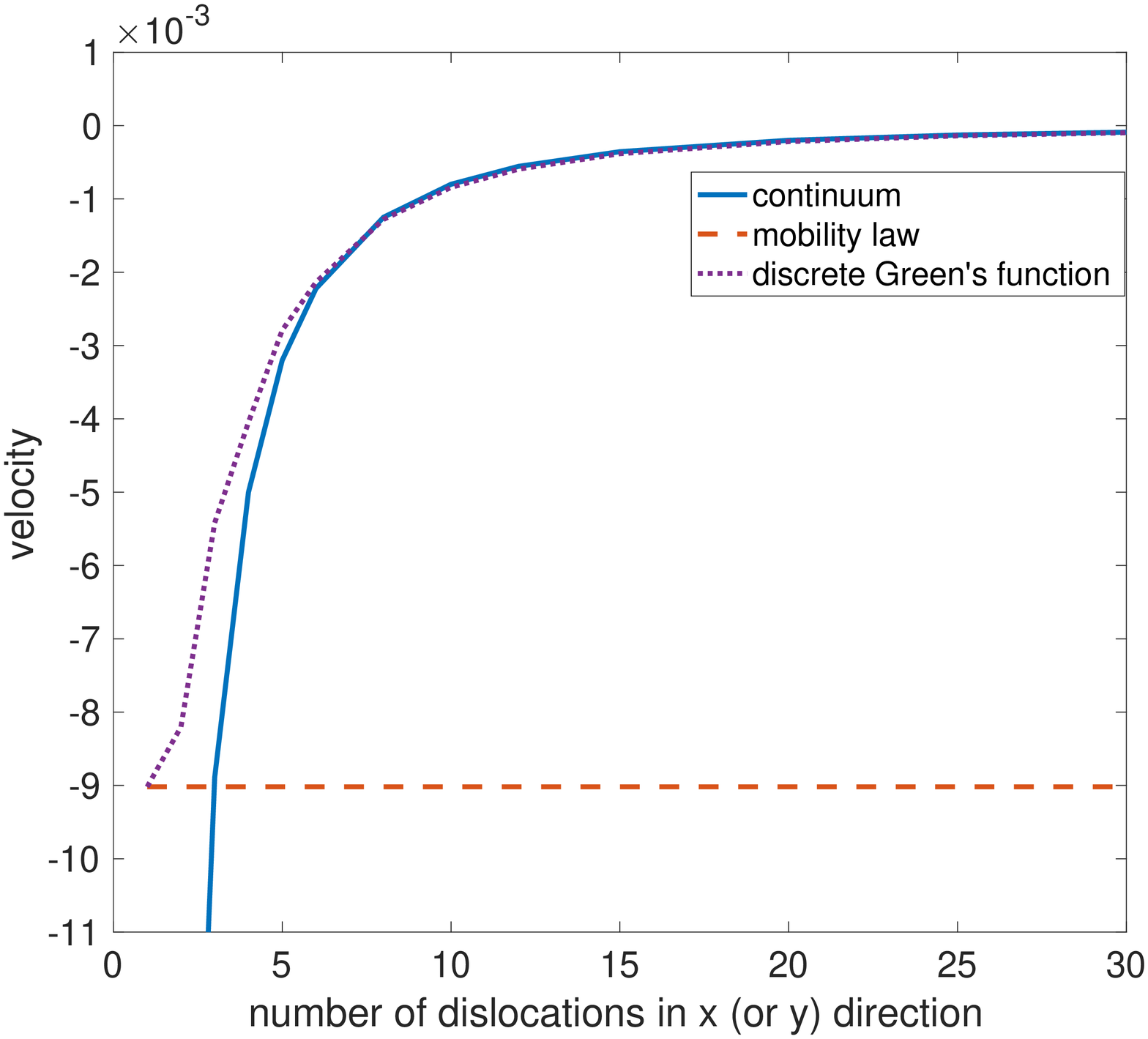}
  \caption{Comparison of  climb velocities for 2D uniform distributed dislocation over $[0,3000b]^2$ with periodic boundary conditions, using our continuum formulation in Eq.~\eqref{eq:v_cl}, the Green's function based discrete dislocation dynamics method in Eq.~\eqref{eqn:discrete0002d}, and the continuum mobility law formulation in Eq.~\eqref{eqn:mobility-law}. The climb velocity unit is $\frac{c_0\mu D_v\Omega}{2\pi (1-\nu)k_BT}$.}
  \label{fig:v_uniform_3m}
\end{figure}

As seen from Figure \ref{fig:v_uniform_3m}, the continuum dislocation climb formulation provides a good approximation for the Green's function discrete dislocation model when the dislocation distribution is not very sparse. This validates our continuum formulation. Whereas the mobility model is good only for a sparse dislocation distribution, otherwise the error is large.
%
%

\subsection{A regular dislocation wall with perturbation in the climb direction}\label{sec:NS1}

Consider the special 2D case discussed in Sec.~\ref{sec:2D}, where the straight edge dislocations are uniformly distributed along the $x$-direction while vary along the $y$-direction, as shown in Fig.~\ref{fig:x_uni}.
It is reduced to a one-dimensional (1D) problem with the variable $y$. Assuming that $c_\infty=c_0$ and $f_{\rm cl}^0=0$, the continuum climb velocity in this case is given by Eq.~\eqref{eq:1D}, which is due to the climb Peach-Koehler force generated by the dislocations.

The dislocations are located at $(x_i,y_j)=(iB, jD-0.02D\sin\frac{4\pi jD}{L_2}-0.02D\cos\frac{2\pi jD}{L_2})$ for integers $i$ and $j$, with $B$ and $D$ being the average inter-dislocation distances in the $x$ and $y$ directions, respectively. We set $B=30b$ and $L_2=3000b$, with different values of $D$.
The dislocation density can be accordingly calculated:
$\rho(x,y)=\rho(y)=\frac{1}{BD(y)}\approx\frac{1}{BD}+\frac{0.08\pi}{BL_2}\cos\frac{4\pi y}{L_2}-\frac{0.04\pi}{BL_2}\sin\frac{2\pi y}{L_2}$, where $D(y)$ is the local inter-dislocation distance in $y$-direction.

Using the continuum climb velocity in Eq.~\eqref{eq:1D}, we have
\begin{equation}
    v(y)=-\frac{2 c_0D_v\mu\Omega}{(1-\nu)k_{\rm B} T}\frac{\rho'(y)}{\rho(y)}\approx\frac{2 c_0D_v\mu\Omega}{(1-\nu)k_{\rm B} T}\cdot
    \frac{0.32\pi^2D\sin\frac{4\pi y}{L_2}+0.08\pi^2D\cos\frac{4\pi y}{L_2}}{L_2^2+0.08\pi DL_2\cos\frac{4\pi y}{L_2}y -0.04\pi DL_2\sin\frac{2\pi y}{L_2}}.\label{eqn:vcl-ex2}
\end{equation}

\begin{figure}[htbp]
\centering
\subfigure[Climb velocity by the continuous model.] {\includegraphics[width = .48\linewidth]{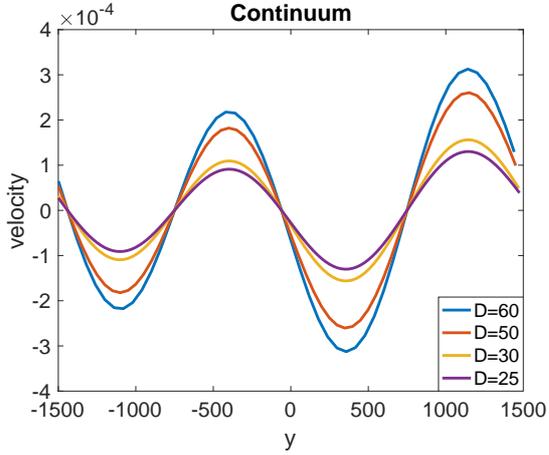} \label{fig:vc}}  \hfill
\subfigure[Climb velocity by discrete Green's function model.] {\includegraphics[width = .48\linewidth]{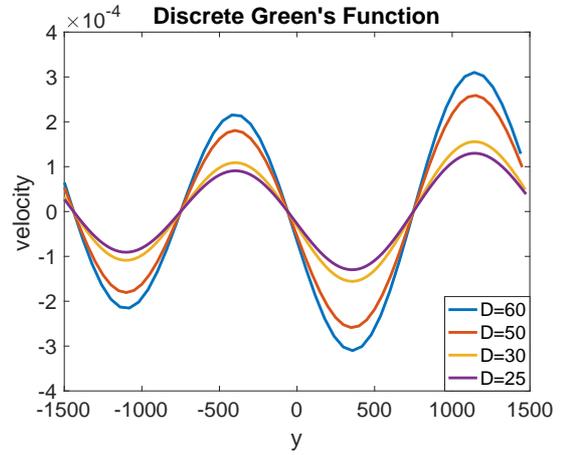}\label{fig:vc_contour}}
\subfigure[Error between results of continuum model in (a) and discrete model in (b).]{\includegraphics[width = .48\linewidth]{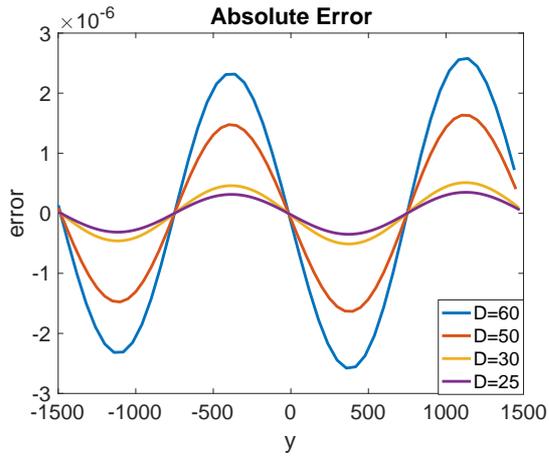}\label{fig:vm}}\hfill
\subfigure[Continuum climb velocity by mobility law.]{\includegraphics[width = .48\linewidth]{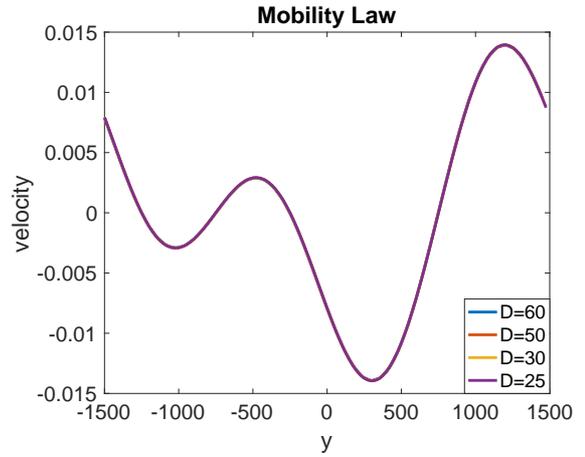}\label{fig:vm_contour}}
\caption{A regular dislocation wall with perturbation in the climb direction: Comparisons of dislocation climb velocities obtained using our continuum formulation in Eq.~\eqref{eq:1D}, the Green's function based discrete dislocation dynamics method in Eq.~\eqref{eqn:discrete0002d}, and the continuum mobility law formulation in Eq.~\eqref{eqn:mobility-law000}. The climb velocity unit is $\frac{c_0\mu D_v\Omega}{2\pi (1-\nu)k_BT}$.}
 \label{fig:x_uni_error}
\end{figure}

We compare the climb velocity obtained using the continuum formulation in Eq.~\eqref{eqn:vcl-ex2}
 with that obtained by using the  discrete Green's function method \cite{gu2015three} given in Eq.~\eqref{eqn:discrete0002d}. Here the Green's function $G(x,y)$ under periodic boundary conditions is calculated by FFT using $\widehat{G}=-\frac{1}{k_1^2+k_2^2}\widehat{\rho^{\rm d}}$, where $\rho^{\rm d}(x,y)=\sum_{i,j}\delta_{r_d}(x-x_i,y-y_j)$ and
  $\delta_{r_d}(x,y)=\frac{1}{2\pi r_d^2}\exp\left(-\frac{x^2+y^2}{2r_d^2}\right)$ is a regularized 2D $\delta$-function.
The climb force $f_{\rm cl}$ in the discrete dislocation model in Eq.~\eqref{eqn:discrete0002d} is calculated  using Eq.~\eqref{eq:f_cl-d} by summing up contributions from dislocations within $20$ periodic images of the domain.

We also compare the climb velocity obtained using mobility law in Eq.~\eqref{eqn:mobility-law000} with those of our continuum formulation in Eq.~\eqref{eqn:vcl-ex2} and the discrete dislocation model in  Eq.~\eqref{eqn:discrete0002d}.
Here the continuum climb force $f_{\rm cl}$ in the mobility law in Eq.~\eqref{eqn:mobility-law000}  is calculated by FFT using Eq.~\eqref{eqn:fft-fcl20}  with $f_{\rm cl}^0=0$. The outer cutoff distance $r_\infty=1616b$, and the dislocation core radius $r_d=2b$.

The comparison results are shown in Fig.~\ref{fig:x_uni_error} for different values of $D$. It can be seen from Fig.~\ref{fig:x_uni_error}(a)-(c) that our continuum climb formulation provide an accurate approximation to the climb velocity of the discrete dislocation model, and converges to the discrete model result as the dislocation distribution becomes dense. The relative errors for these values of $D$ ($50b$ down to $25b$) are less than the order of $1\%$. Whereas the continuum climb velocity based on mobility law shown in Fig.~\ref{fig:x_uni_error}(d) is not able to give correct results with respect to those of the discrete model, with results  $O(10^2)$ larger than those of the discrete model (and our continuum formulation) and  different profile as a function of $y$.


\subsection{A regular dislocation wall with perturbation in both the Burgers vector and the climb directions}

\begin{figure}[ht]
\centering
  \includegraphics[width=0.5\textwidth]{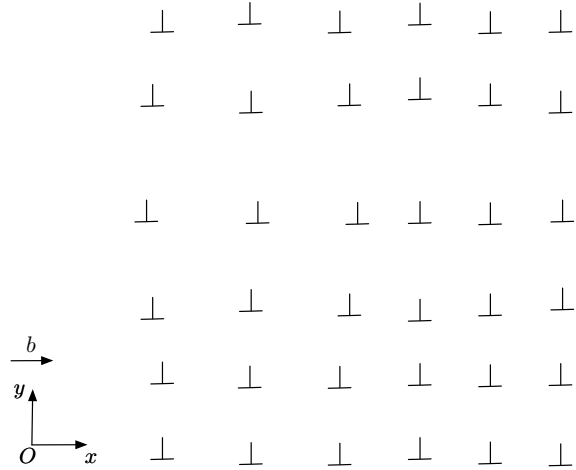}
  \caption{A regular dislocation wall with perturbation in both the Burgers vector and the climb directions.}
  \label{fig:posi_dis}
\end{figure}

Consider a dislocation wall where the straight edge dislocations are subject to perturbations both in the Burgers vector direction (i.e. the $x$-direction) and the
climb direction  (i.e. the $y$-direction) and vary along the $y$-direction, as illustrated in Fig.~\ref{fig:posi_dis}.
Assume $c_\infty=c_0$ and $f_{\rm cl}^0=0$, the continuum climb velocity  is given by  Eq.~\eqref{eq:v_cl}, with only the first integral term which is due to the climb Peach-Koehler force generated by the dislocations.

We consider the simulation domain $[0,L_1]\times[0,L_2]$ with periodic boundary conditions.
The dislocations are located at  $$(x_i,y_j)=\left(iB-0.02B\sin\frac{4\pi iB}{L_1}\sin\frac{10\pi jD}{L_2}, jD-0.02D\sin\frac{20\pi iB}{L_1}\sin\frac{4\pi jD}{L_2}\right)$$ for integers $i$ and $j$, with $B$ and $D$ being the average inter-dislocation distances in the $x$ and $y$ directions, respectively. We set $D=50b$, $B=30b$, $L_1=40B=1200b$, and $L_2=60D=3000b$.
The dislocation density can be accordingly calculated:
\begin{flalign}
\rho(x,y)=&\frac{1}{B(x)D(y)}\nonumber\\
\approx&\frac{1}{BD}+
\frac{0.2\pi }{BL_2}\sin\frac{4x\pi}{L_1}\cos\frac{10\pi y}{L_2}+\frac{0.4\pi }{DL_1}\cos\frac{20\pi x}{L_1}\sin\frac{4\pi y}{L_2}\nonumber\\
&+\frac{0.08\pi^2 }{L_1L_2}\sin\frac{4\pi x}{L_1}\sin\frac{4\pi y}{L_2}\cos\frac{20\pi x}{L_1}\cos\frac{10\pi y}{L_2}\nonumber\\
&-\frac{0.0064\pi^2 }{L_1L_2}\sin\frac{20\pi x}{L_1}\sin\frac{10\pi y}{L_2}\cos\frac{4\pi x}{L_1}\cos\frac{4\pi y}{L_2},
\end{flalign}
where $B(x)$ and $D(y)$ are the local inter-dislocation distances in the $x$- and $y$-directions, respectively.

\begin{figure}[htbp]
\centering
\subfigure[] {\includegraphics[width = .48\linewidth]{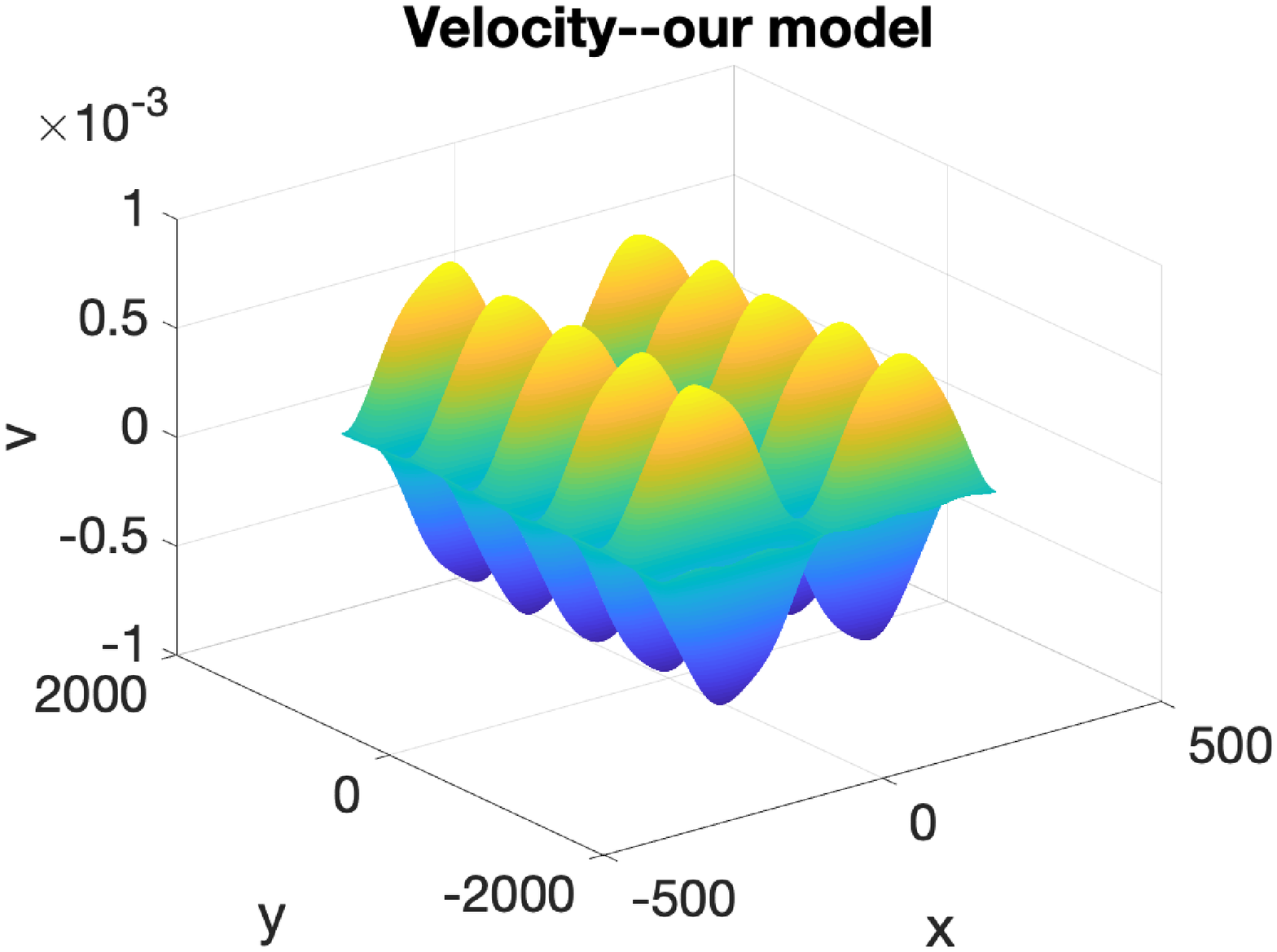} \label{fig:vc}}  \hfill
\subfigure[] {\includegraphics[width = .48\linewidth]{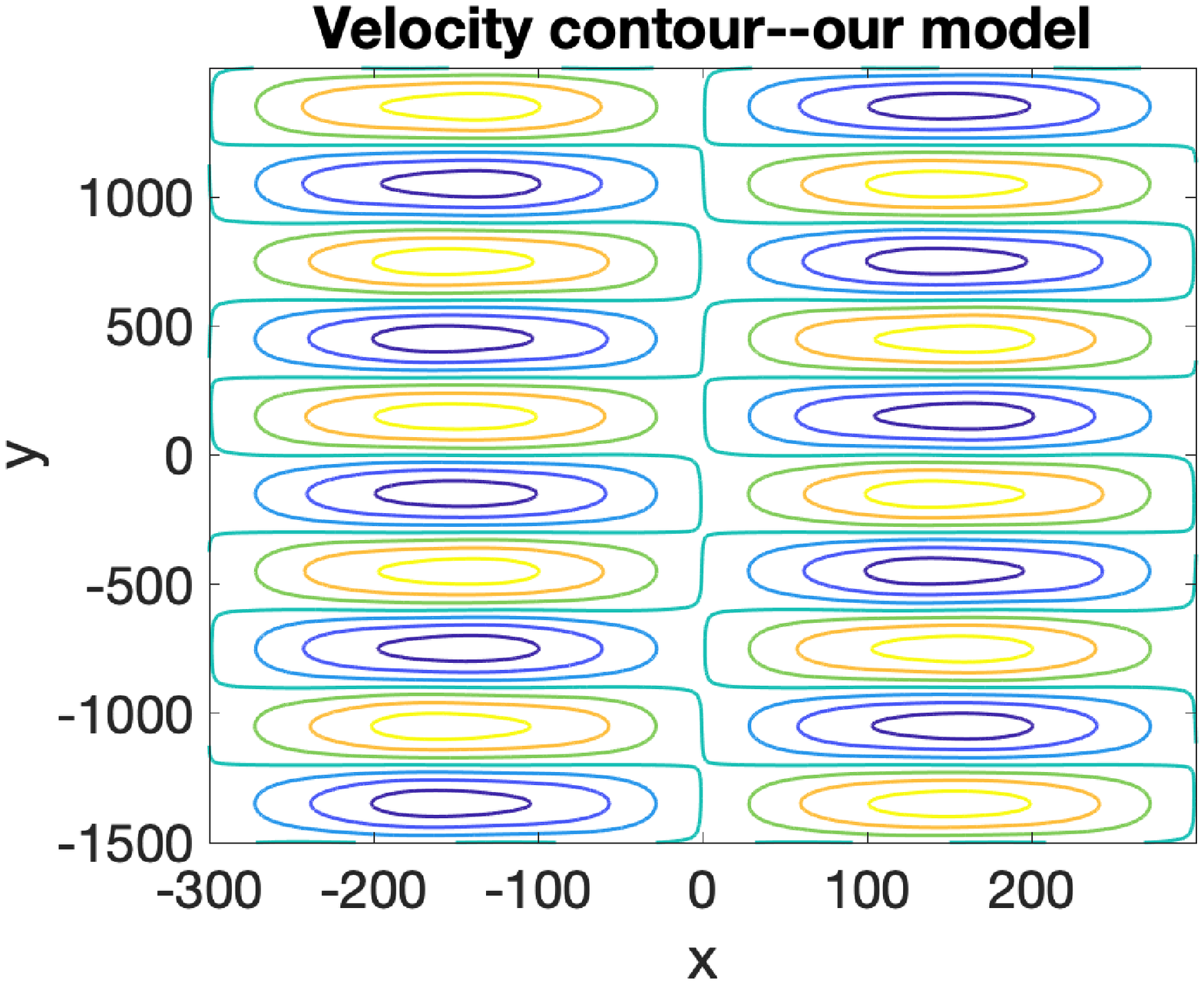}\label{fig:vc_contour}}
\subfigure[]{\includegraphics[width = .48\linewidth]{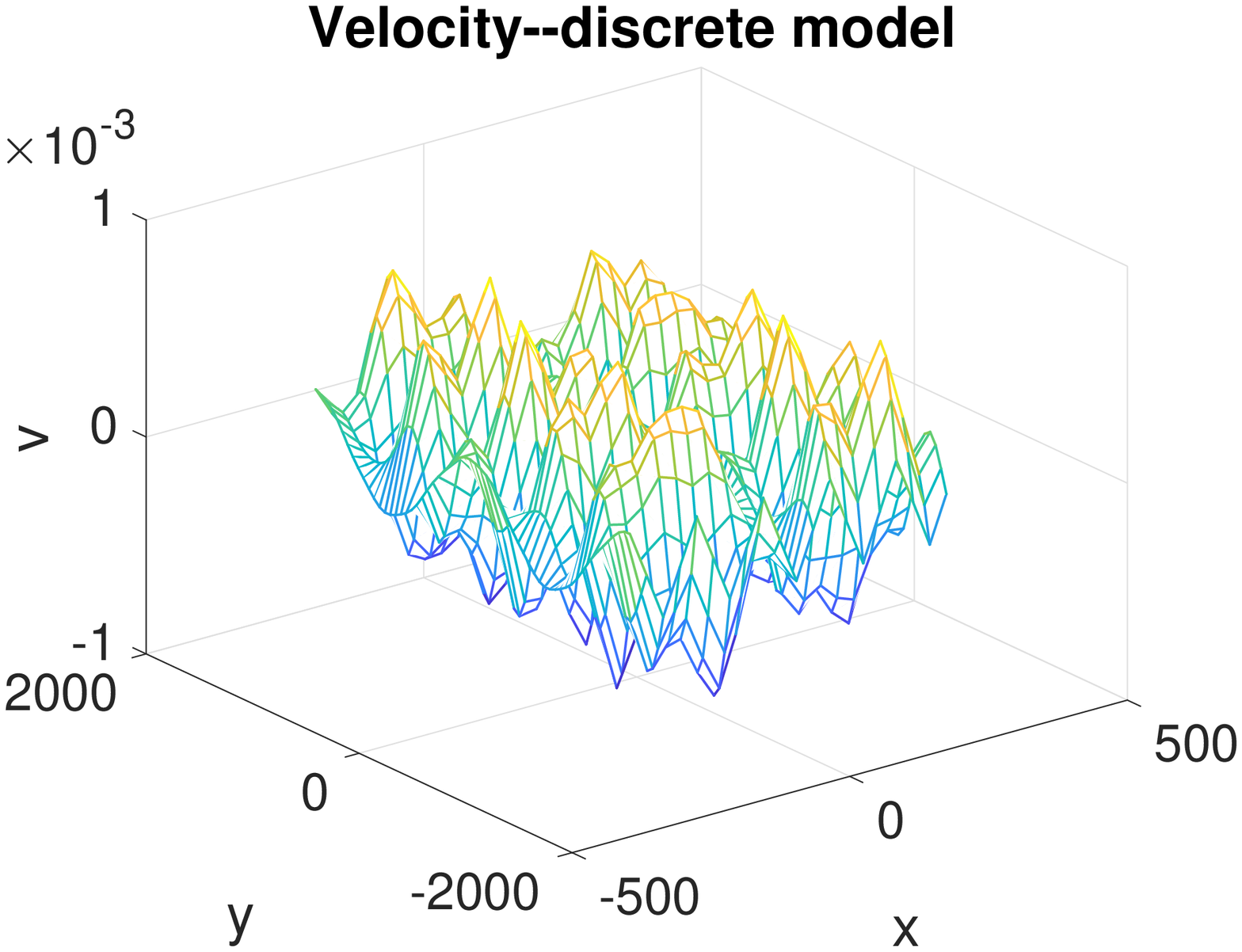} \label{fig:vd}}\hfill
\subfigure[]{\includegraphics[width = .48\linewidth]{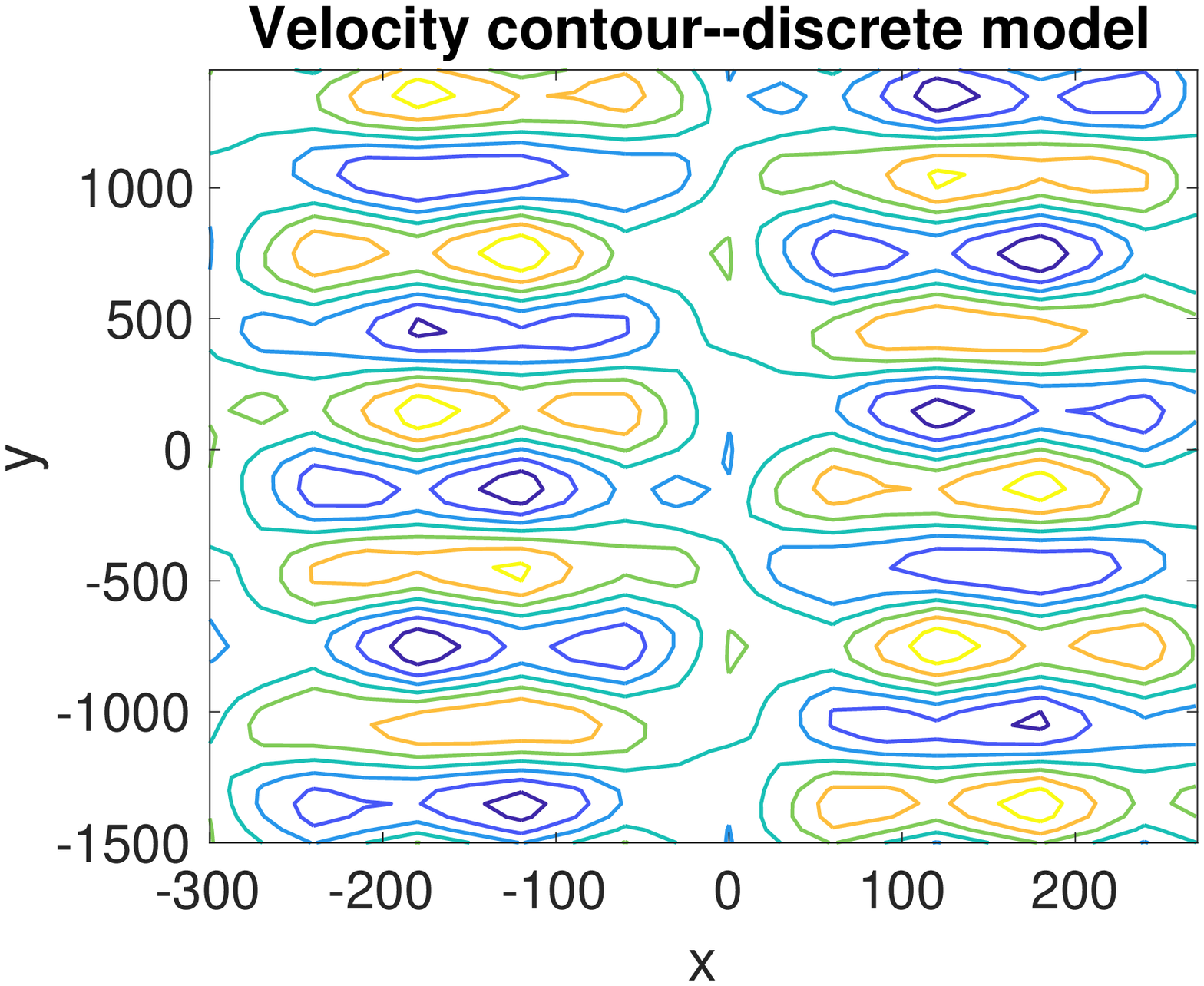}\label{fig:vd_contour}}
\subfigure[]{\includegraphics[width = .48\linewidth]{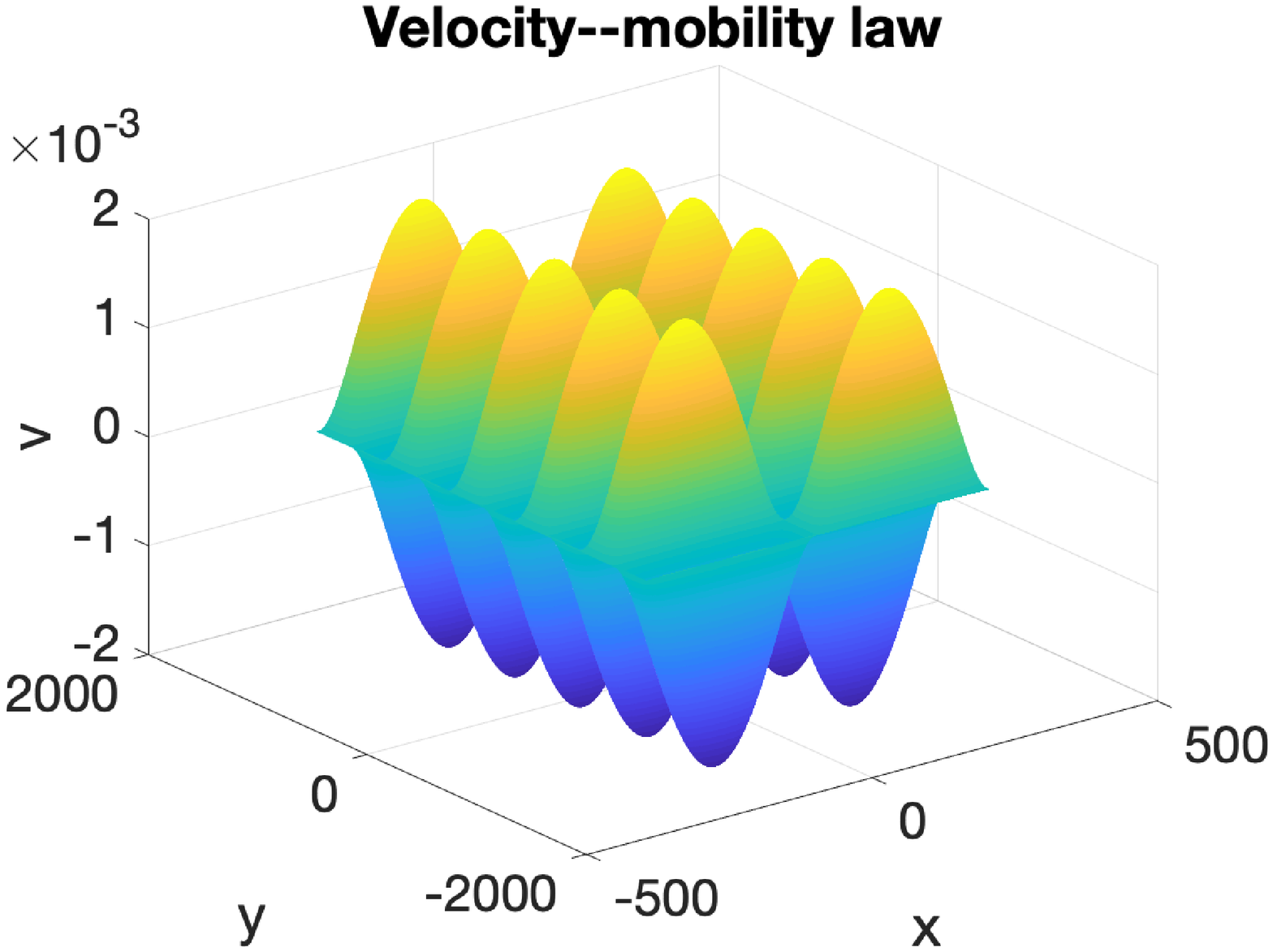}\label{fig:vm}}\hfill
\subfigure[]{\includegraphics[width = .48\linewidth]{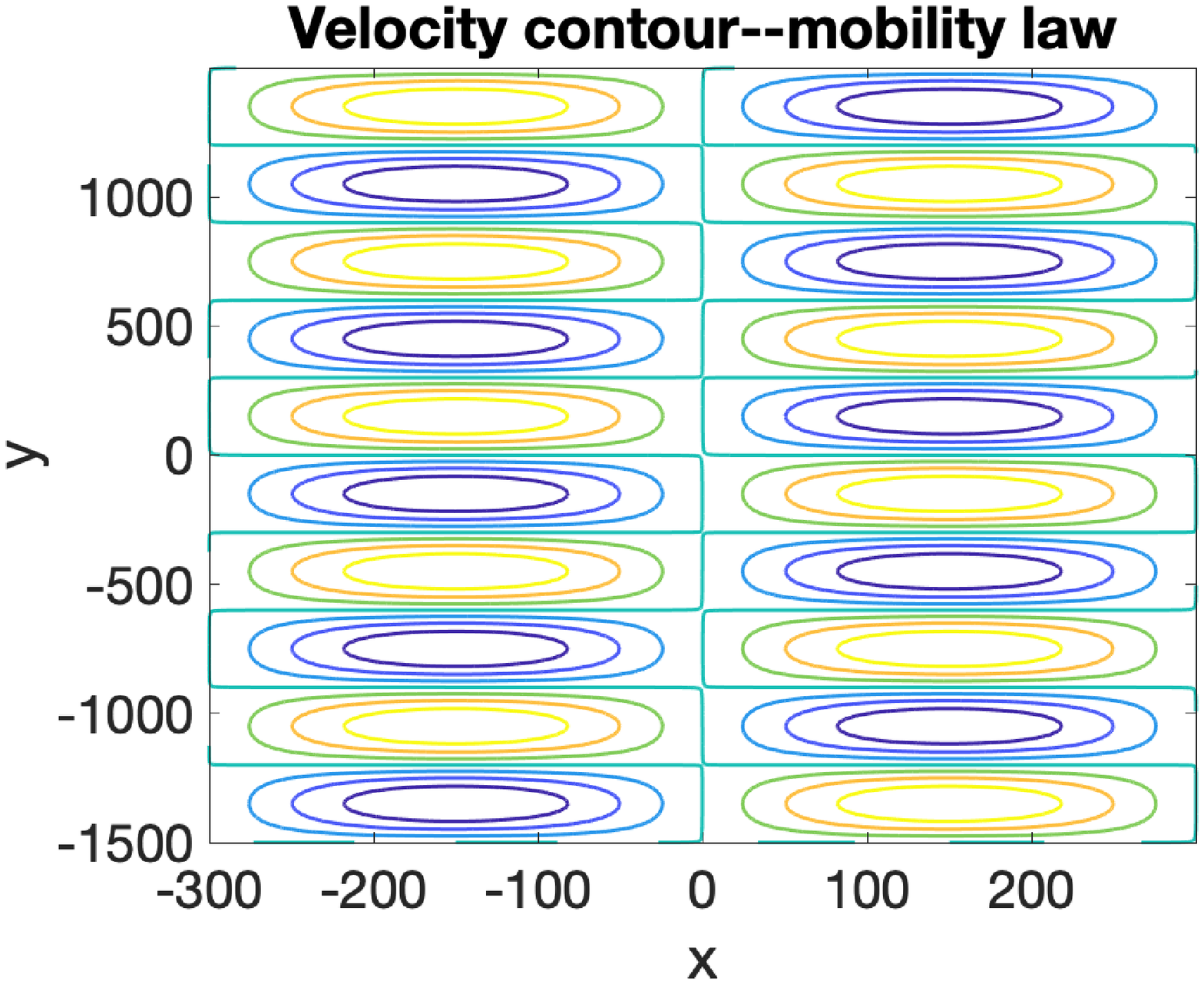}\label{fig:vm_contour}}
\caption{A regular dislocation wall with perturbation in both the Burgers vector and the climb directions: Comparisons of dislocation climb velocities obtained using our continuum formulation in Eq.~\eqref{eq:v_cl}, the Green's function based discrete dislocation dynamics method in Eq.~\eqref{eqn:discrete0002d}, and the continuum mobility law formulation in Eq.~\eqref{eqn:mobility-law}. The climb velocity unit is $\frac{c_0\mu D_v\Omega}{2\pi (1-\nu)k_BT}$.}
\label{fig:ex3}
\end{figure}


\begin{figure}[htbp]
\centering
\subfigure[]{\includegraphics[width = .48\linewidth]{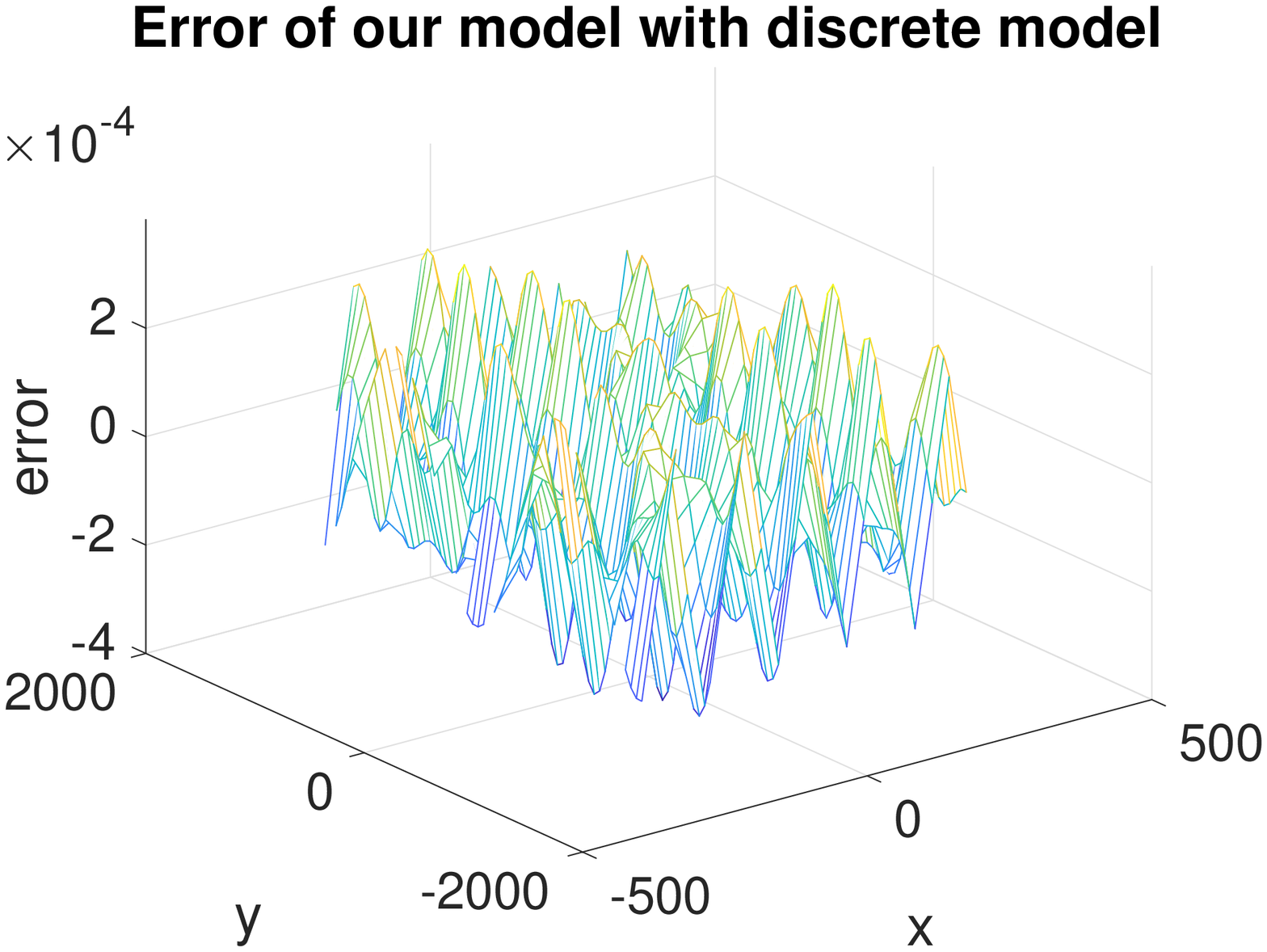}\label{fig:err_vc_vd}}\hfill
\subfigure[]{\includegraphics[width = .48\linewidth]{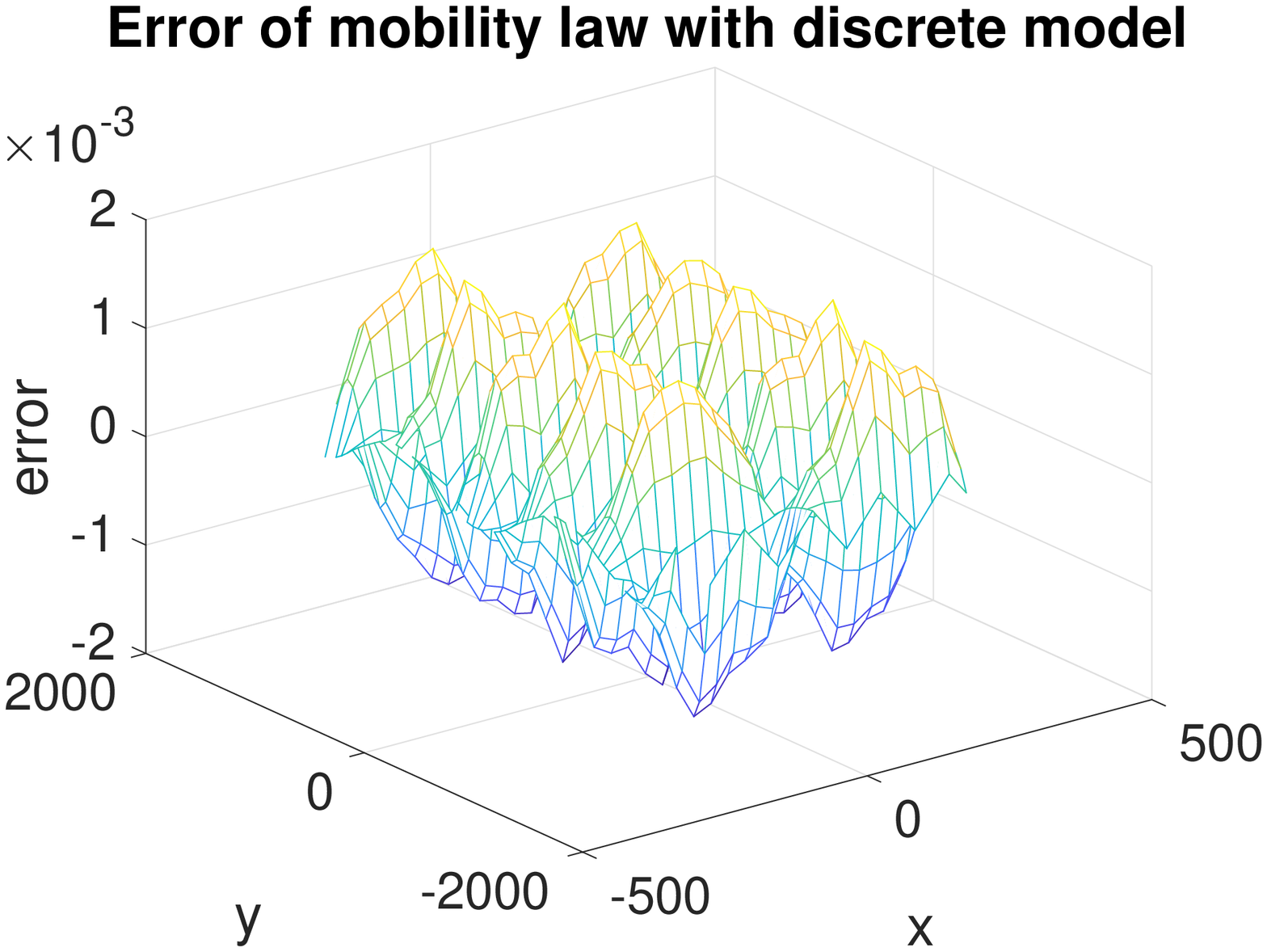}\label{fig:err_vm_vd}}
\caption{Errors of our continuum climb velocity formulation and the continuum mobility law model with respect to the discrete dislocation climb model, based on the results in Fig.~\ref{fig:ex3}. The climb velocity unit is $\frac{c_0\mu D_v\Omega}{2\pi (1-\nu)k_BT}$.}
\label{fig:ex3-errors}
\end{figure}

The continuum dislocation climb velocity formulation is given in Eq.~\eqref{eq:v_cl} with $c_\infty=c_0$ and $f_{\rm cl}=0$, and we calculate it using FFT by Eq.~\eqref{eq:sol}.
We compare the climb velocity obtained using this continuum formulation
 with that obtained by using the  discrete Green's function method \cite{gu2015three} given in Eq.~\eqref{eqn:discrete0002d}, which is calculated the same way as in the previous example.
We also compare the climb velocity obtained using mobility law in Eq.~\eqref{eqn:mobility-law} with the above two results.
The continuum climb force $f_{\rm cl}$ in Eq.~\eqref{eq:f_cl} needed in the mobility law is calculated by FFT using Eq.~\eqref{eqn:fft-fcl20}  with $f_{\rm cl}^0=0$. The outer cutoff distance $r_\infty=1616b$, and the dislocation core radius $r_d=2b$.

Comparison results of the three models are shown in Figs.~\ref{fig:ex3} and \ref{fig:ex3-errors}.
It can be seen that our continuum climb formulation gives an accurate approximation to the climb velocity of the discrete dislocation model,
whereas  the error of the continuum  mobility law model as an approximation to the discrete model is large, which is of order of $100\%$.

\section{Incorporation of Climb in Dislocation Density based Dislocation Dynamics Models}

Our continuum dislocation climb formulations (Eq.~\eqref{eq:v_cl} for 2D and Eq.~\eqref{eqn:vcl-3d} for 3D) can be incorporated in any
 dislocation density based continuum dislocation dynamics frameworks.
In this section, we present incorporation of this continuum formulation in our continuum dislocation dynamics model based on dislocation density potential functions (DDPFs) \cite{zhuyichao2015continuum,Niu2018,zhu2014continuum,xiang2009continuum,zhu2010continuum}.

In the DDPF framework of continuum dislocation dynamics framework, the dislocations with the same Burgers vector $\mathbf b$ are represented by a pair of scalar functions (DDPFs) $\phi(x,y,z)$ and $\psi(x,y,z)$, such that the intersection of the contour lines determined by $\phi(x,y,z)=ib$ and $\psi(x,y,z)=jb$, where $b$ is the length of the Burgers vector and  $i$ and $j$ are integers, are the dislocation lines. The advantages of this representation include its simple representation of distributions of curved dislocations, and the
connectivity condition of dislocations is automatically satisfied. Moreover, geometric quantities of dislocations, such as the local dislocation line direction and dislocation curvature, can be easily calculated from the two DDPFs $\phi$ and $\psi$.

Especially, the local dislocation line direction is
\begin{equation}
\pmb \xi=\frac{\nabla \phi\times \nabla \psi}{\|\nabla \phi\times \nabla \psi\|},
\end{equation}
and dislocation density is
\begin{equation}
\rho=\frac{1}{b^2}\|\nabla \phi\times \nabla \psi\|.
\end{equation}

Dynamics of the dislocation density is given by
\begin{flalign}
\phi_t+\mathbf v\cdot\nabla  \phi&=0, \label{eqn:phi-evolution}\\
\psi_t+\mathbf v\cdot\nabla  \psi&=0,\label{eqn:psi-evolution}
\end{flalign}
where $\mathbf v$ is the local dislocation velocity including glide velocity and climb velocity. Here we focus on the climb motion of dislocations, and we have
\begin{equation}
\mathbf v=\mathbf v_{\rm cl}=v_{\rm cl}(\pmb \xi\times \mathbf b/b),\label{eqn:ex3-vcl}
\end{equation}
where the continuum climb velocity $v_{\rm cl}$  is given by  Eq.~\eqref{eq:v_cl} in 2D or Eq.~\eqref{eqn:vcl-3d} in 3D.

\begin{figure}[htbp]
\centering
\subfigure[] {\includegraphics[width = .32\linewidth]{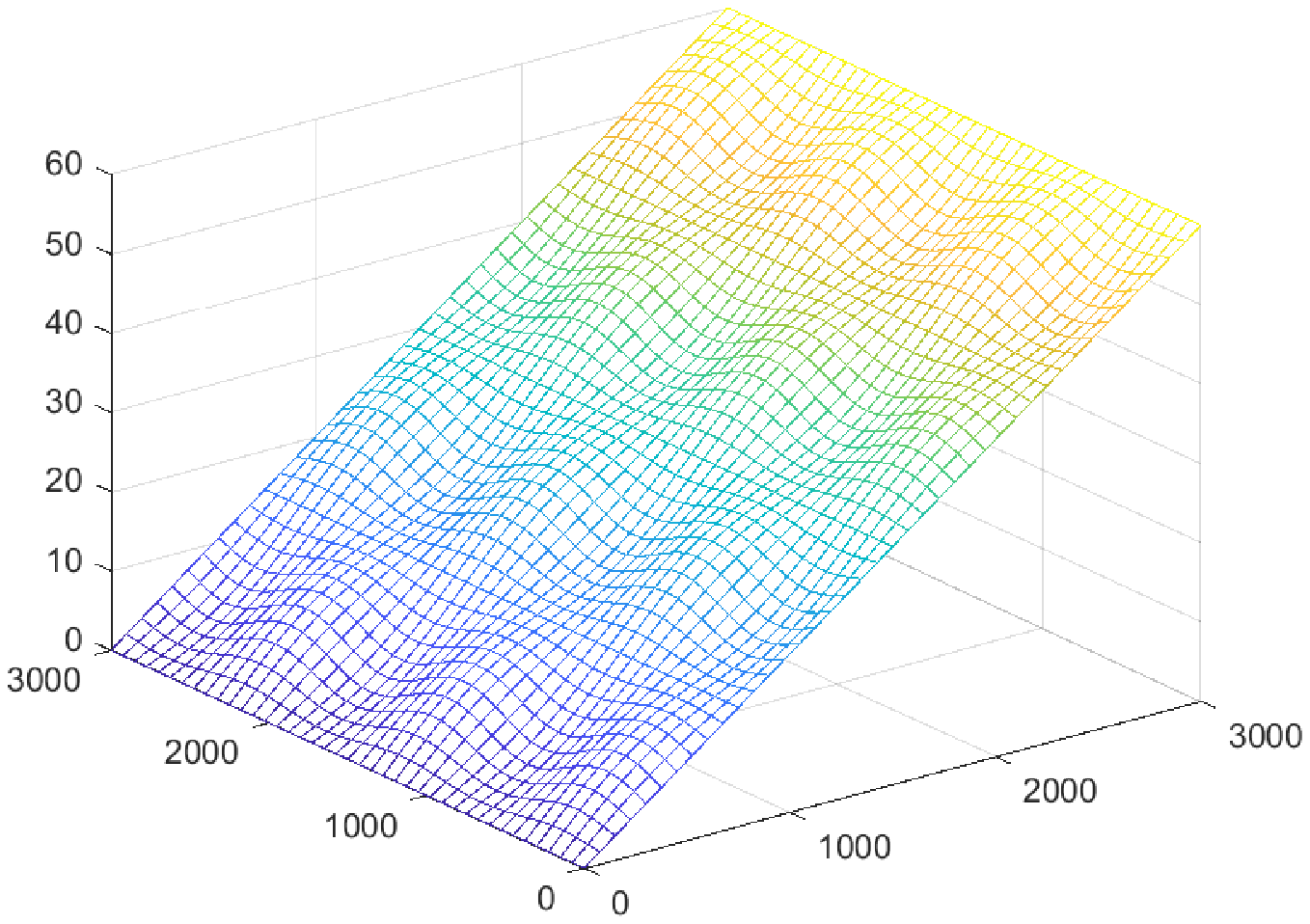}}
\subfigure[] {\includegraphics[width = .25\linewidth]{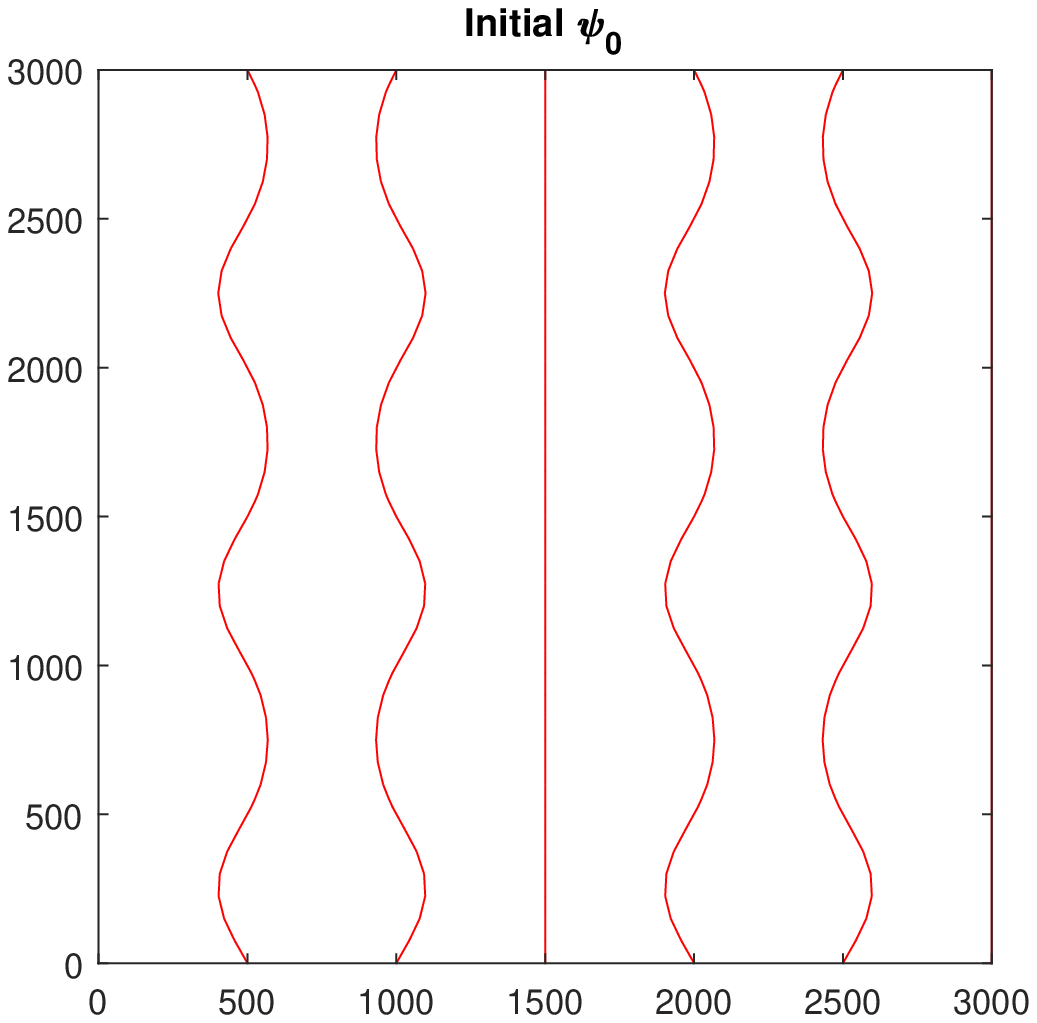}}\\
\subfigure[]{\includegraphics[width = .32\linewidth]{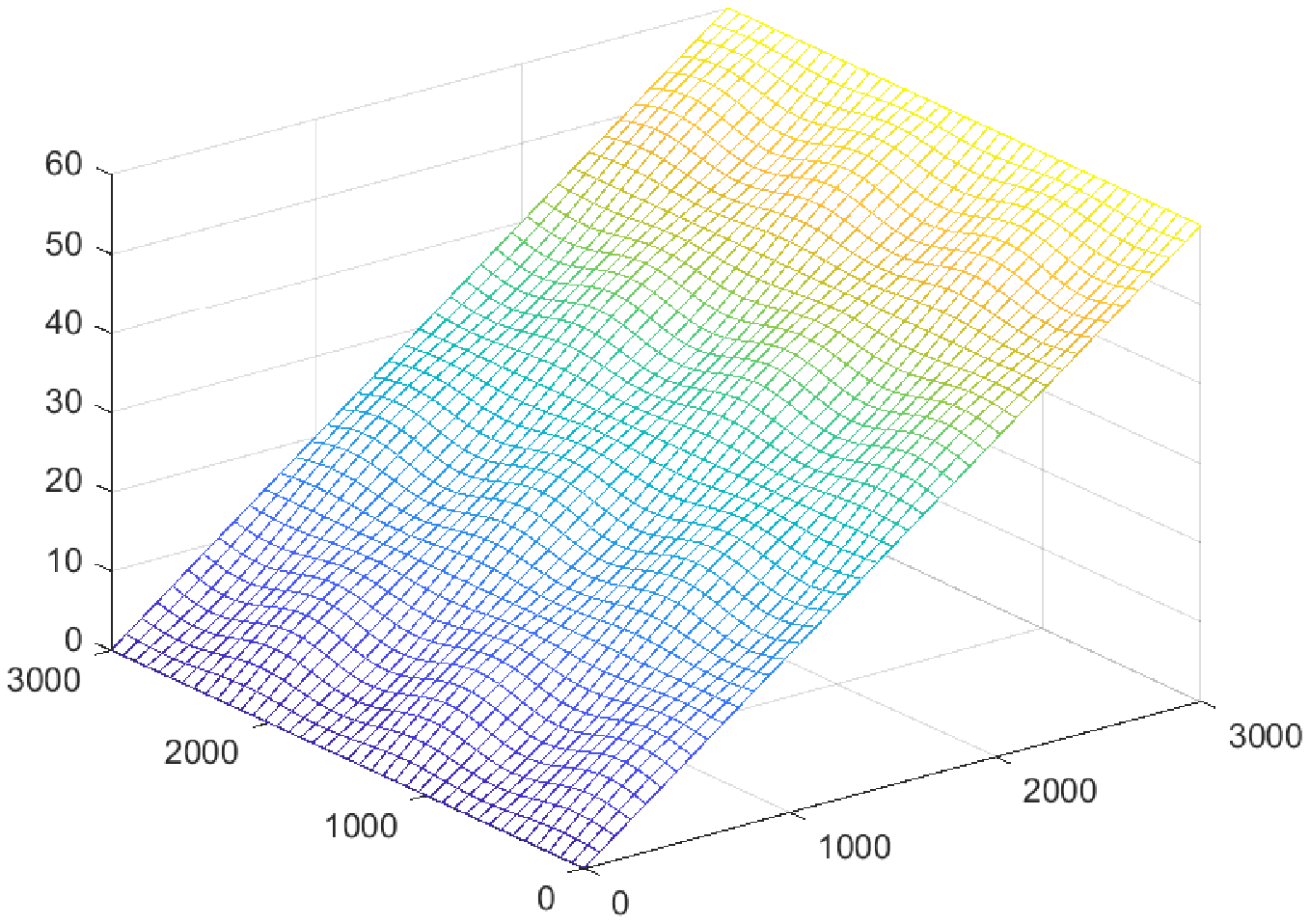}}
\subfigure[]{\includegraphics[width = .25\linewidth]{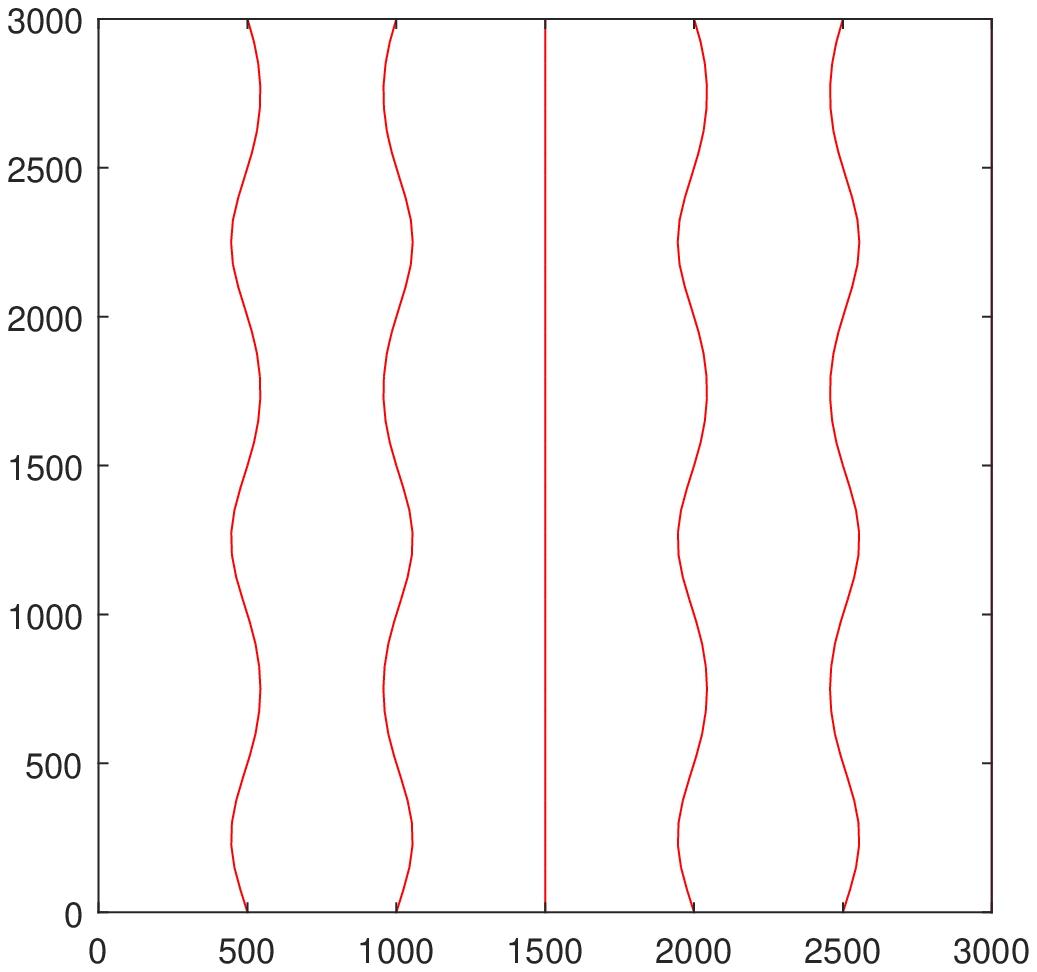}}\\
\subfigure[]{\includegraphics[width = .32\linewidth]{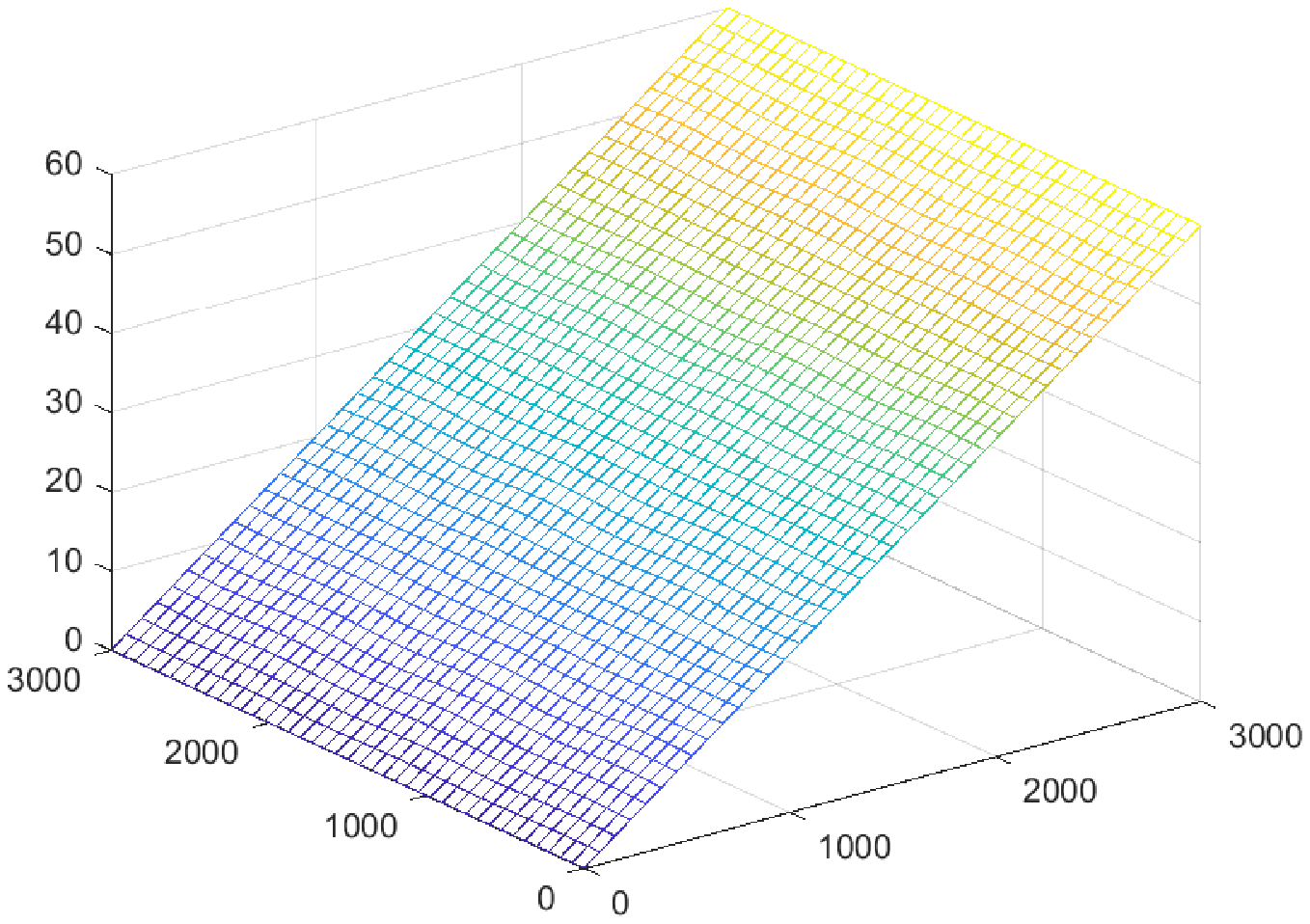}}
\subfigure[]{\includegraphics[width = .25\linewidth]{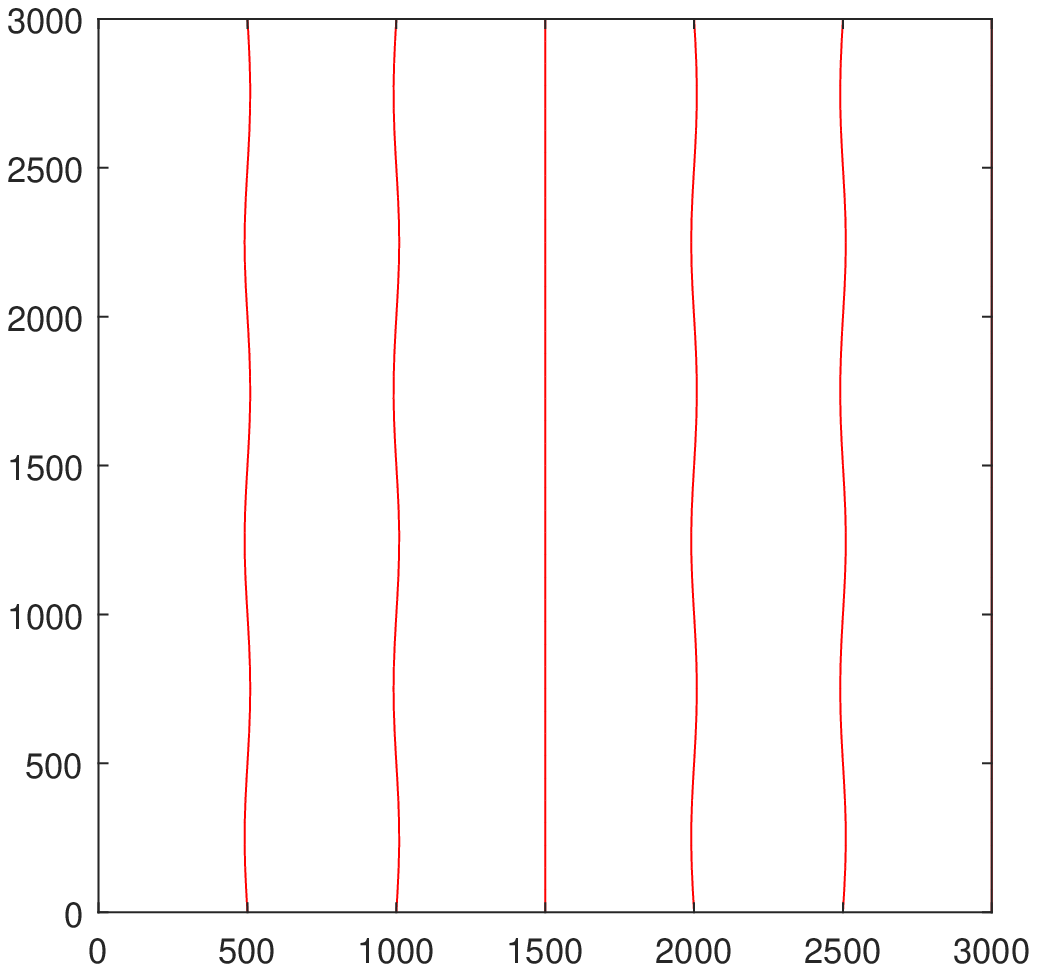}}\\
\subfigure[]{\includegraphics[width = .32\linewidth]{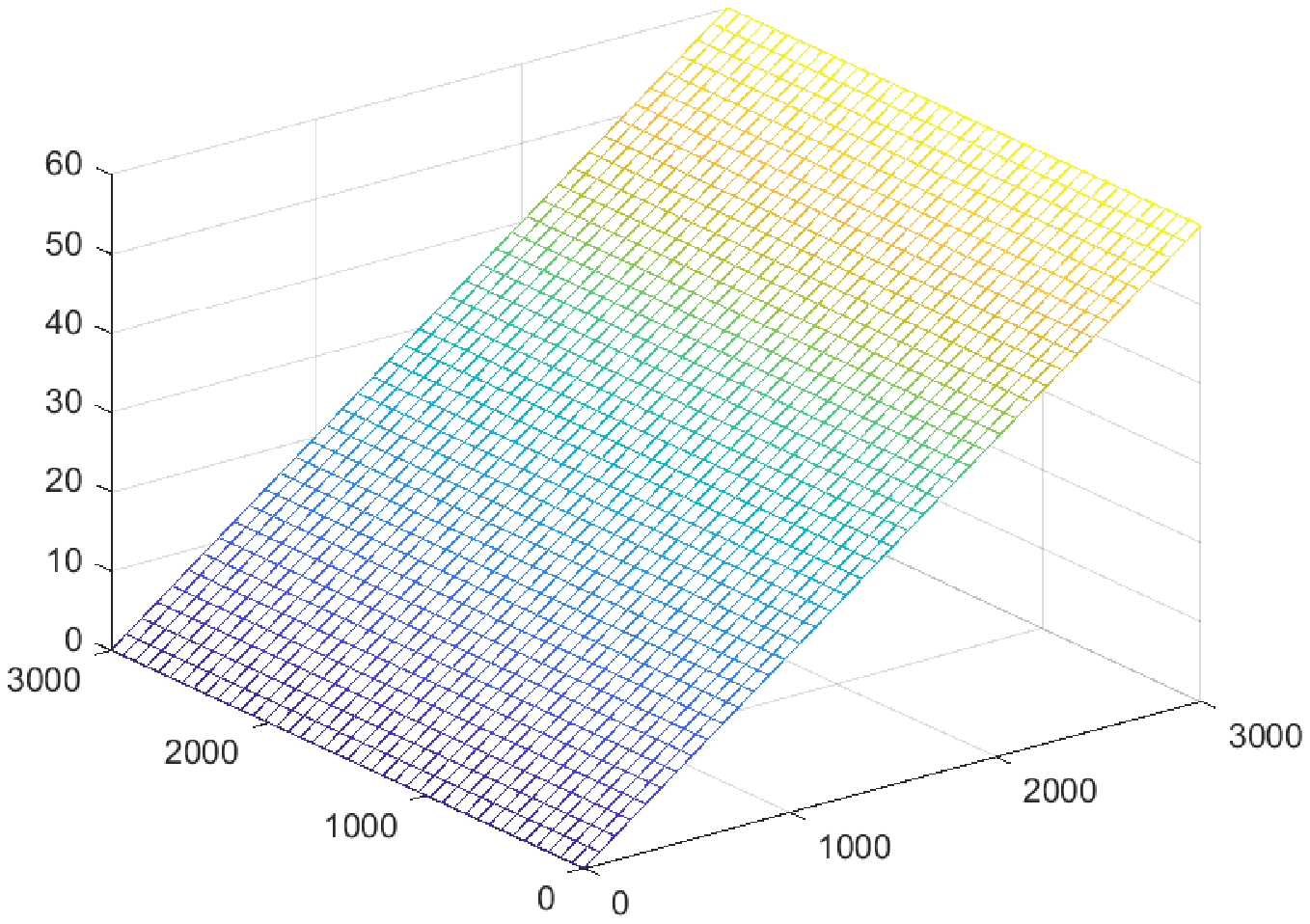}}
\subfigure[]{\includegraphics[width = .25\linewidth]{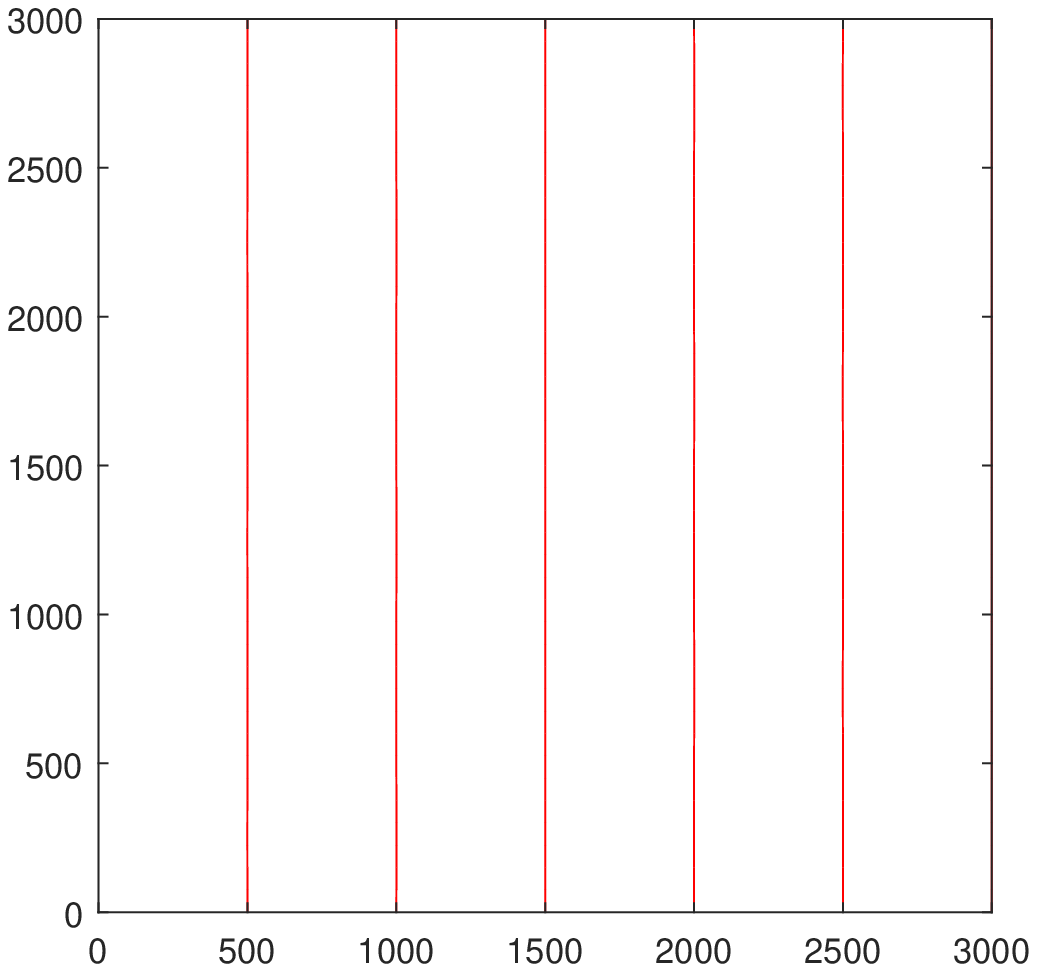}}
\caption{Evolution of distribution of perturbed straight dislocation array by using the obtained continuum climb velocity in the continuum dislocation dynamics framework based on DDPF. Snapshots at four time steps are shown: from the initial distribution on the top to the converged distribution of straight dislocations at the bottom. The left panel of images show the profile of the DDPF $\psi$, and the left panel of images show the profiles of some selected dislocations. The length unit is $b$.}
\label{fig:dyn1}
\end{figure}

\begin{figure}[htbp]
\centering
\subfigure[$\rho$] {\includegraphics[width = .35\linewidth]{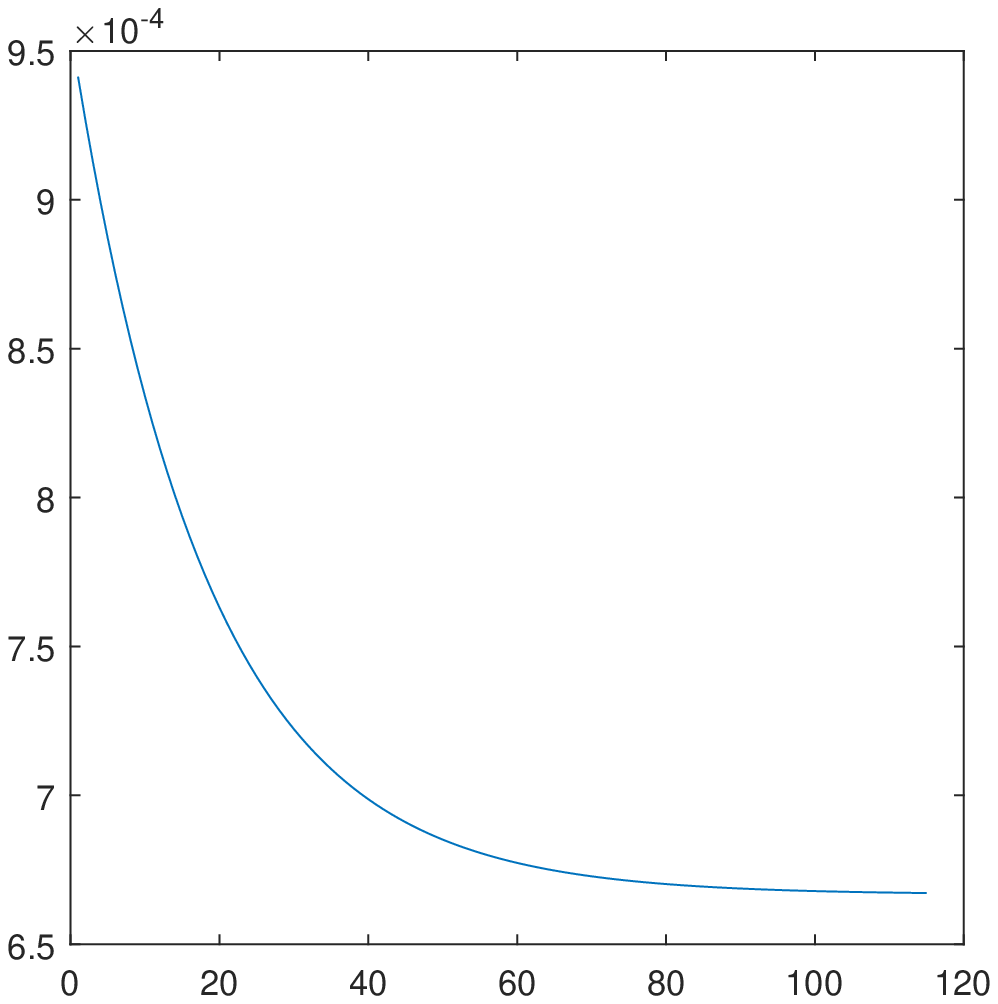}}
\subfigure[$\log(\rho-\rho_0)$ with $\rho_0=\frac{1}{BD}$] {\includegraphics[width = .35\linewidth]{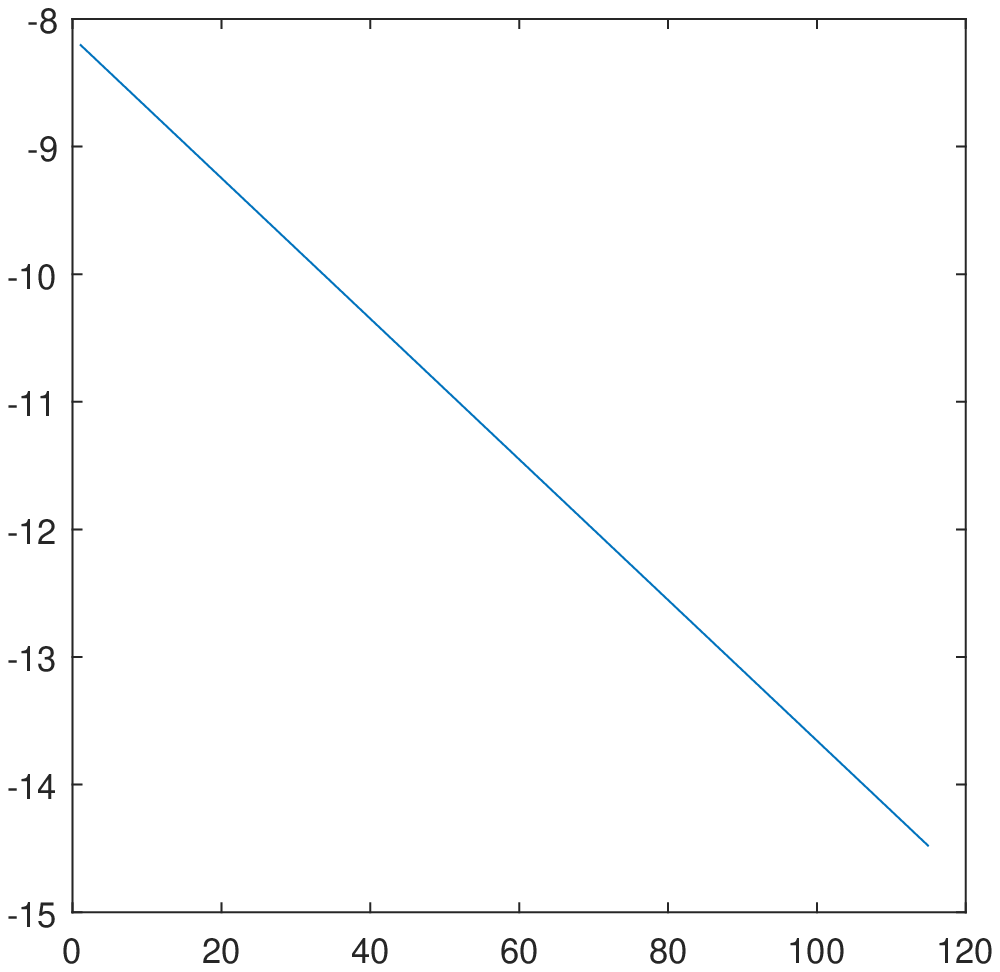}}\\
\caption{Evolution of distribution of perturbed straight dislocation array by using the obtained continuum climb velocity in the continuum dislocation dynamics framework based on DDPF: Evolution of the dislocation density $\rho$ at a fixed point. The unit of $\rho$ is $1/b^2$. The horizontal axis is time with unit $0.5\times10^{-3}\left(\frac{c_0\mu D_v\Omega}{(1-\nu)b^2k_BT}\right)^{-1}$.}
\label{fig:dyn2}
\end{figure}

We illustrate the implementation of this continuum dislocation dynamics model by a simple example.
Consider a simulation domain of size $L_1\times L_2\times L_3$, with $D=50b$, $B=30b$, $L_1=40B=1200b$, $L_2=L_3=60D=3000b$. The Burgers vector is $\mathbf b=(b,0,0)$. Periodic boundary conditions are used for dislocation distributions, i.e. for $\rho$ and $\pmb \xi$.
The initial dislocation distribution is given by
\begin{flalign}
\phi(x,y,z)=&\frac{b}{B}x,\\
\psi(x,y,z)=&\frac{b}{D}z+2b\sin\frac{6\pi y}{L_2}\sin\frac{4\pi z}{L_3}.
\end{flalign}
Here for a periodic dislocation distribution, $\phi$ and $\psi$ are periodic after subtracting the linear functions.
In this case, the dislocation climb velocity in Eq.~\eqref{eqn:ex3-vcl} is in the form
$\mathbf v_{\rm cl}=(0,v_y,v_z)$,
and the evolution equation in Eq.~\eqref{eqn:phi-evolution} is reduced to $\phi_t=0$. Thus we only need to solve Eq.~\eqref{eqn:psi-evolution} for the evolution of the distribution of dislocations, which is reduced to an evolution in the $yz$ plane.

Simulation result of evolution of the dislocation distribution is shown in Figs.~\ref{fig:dyn1} and \ref{fig:dyn2}. The dislocations are becoming straight during the evolution under the Peach-Koehler climb force and vacancy diffusion. The perturbation in the dislocation profile decays exponentially with time, as shown in Fig.~\ref{fig:dyn2}. The slop of $\log(\rho-\rho_0)$ at a point, where $\rho_0=\frac{1}{BD}$ is the converged, unperturbed density of dislocations, as shown in Fig.~\ref{fig:dyn2}(b), is about $1.09\times10^{-4}\frac{c_0\mu D_v\Omega}{(1-\nu)b^2k_BT}$. In fact, this is the special 3D case discussed at the end of Sec.~\ref{subsection:3D}, whose continuum climb velocity is given in Eq.~\eqref{eq:3D111}. Using Eq.~\eqref{eq:3D111}, it can be calculated that this decay rate is
$$b^2\left( k_2^2+k_3^2\right)\frac{2c_0\mu D_v\Omega}{(1-\nu)b^2k_BT}=1.14\times 10^{-4}\frac{c_0\mu D_v\Omega}{(1-\nu)b^2k_BT},$$
where $k_2=\frac{6\pi}{L_2}$ and $k_3=\frac{4\pi}{L_3}$.
The simulation result agrees perfectly with the linear stability analysis result.

\section{Discussion}

Self-climb of dislocations by vacancy pipe diffusion also plays an important role in irradiated materials~\cite{anderson2017theory,Kroupa,dudarev2013}.
We generalize the continuum dislocation climb velocity to include dislocation self-climb as follows,
which is based on the discrete dislocation dynamics model for pipe diffusion-assisted self-climb~\cite{Niu2017,Niu2019,Niu2021}:
\begin{align}
v_{\rm cl}^p=D_v^p\frac{\ud^2 c_{d}^p}{\ud s^2},
\end{align}
where $D_v^p$ is the pipe diffusion coefficient, $c_0^p$ is the equilibrium vacancy concentration around the dislocation  $c_{d}^p=c_0^p\exp(-\frac{f_{\rm cl}(s)\Omega}{bk_{\rm B}T})$, and $s$ is the arc length function along the dislocation line.

In a dislocation density based continuum model, for an arbitrary smooth function $f(x,y,z)$, along the dislocation line, we have
\begin{align}
\frac{\ud f}{\ud s}=\frac{\partial f}{\partial x}\frac{\partial x}{\partial s}+\frac{\partial f}{\partial y}\frac{\partial y}{\partial s}+\frac{\partial f}{\partial z}\frac{\partial z}{\partial s} = \nabla f\cdot \bm{\xi},
\end{align}
where $\bm{\xi}$ is the local dislocation line direction.
Therefore, under the continuum framework, we have $\frac{\ud c_d^p}{\ud s}=\nabla c_d^p\cdot \bm{\xi}$, and
\begin{align}
\frac{\ud^2 c_d^p}{\ud s^2}=&\frac{\ud (\nabla c_d^p\cdot \bm{\xi})}{\ud s}=\nabla(\nabla c_d^p\cdot \bm{\xi})\cdot\bm{\xi}.
\end{align}
With the assumption that $f_{\rm cl}\ll bk_{\rm B}T/\Omega,$ $c_d^p\approx c_0^p\left(1-\frac{f_{\rm cl}\Omega}{bk_{\rm B}T}\right),$ which leads to the continuum self climb velocity due to pipe diffusion
\begin{align}
    v_{\rm cl}^p=-\frac{D_v^pc_0^p\Omega}{bk_{\rm B}T}\nabla(\nabla f_{\rm cl}\cdot \bm{\xi})\cdot\bm{\xi}.\label{eqn:pipe-v-cl}
\end{align}

With the vacancy pipe diffusion assisted self-climb, the continuum total climb velocity is
\begin{equation}
v_{\rm cl}^{\rm tot}=v_{\rm cl}+v_{\rm cl}^p,
\end{equation}
where $v_{\rm cl}$ is the continuum climb velocity due to vacancy bulk diffusion obtained in Sec.~\ref{sec:2} (Eq.~\eqref{eq:v_cl} for two dimensional problems and Eq.~\eqref{eqn:vcl-3d} for three dimensions), and $v_{\rm cl}^p$ is the continuum self climb velocity due to pipe diffusion given above in Eq.~\eqref{eqn:pipe-v-cl}.


\section{Conclusions}

In this paper, we have derived a continuum formulation for dislocation climb velocity based on densities of dislocations. The obtained continuum formulation is an approximation of the Green's function based discrete dislocation dynamics formulation in \cite{gu2015three}. This continuum climb velocity formulation provides a good approximation for a not very sparse distribution of dislocations, whereas the continuum climb formulation based on mobility law is valid only in the sparse limit of dislocation distributions, as examined with the discrete dislocation dynamics model.

The continuum dislocation climb formulation has the advantage of accounting for both the long-range effect of vacancy bulk diffusion and that of the Peach-Koehler climb force, and the two long-range effects are canceled into a short-range effect (i.e., an integral that has fast-decaying kernel and can be calculated by truncation over some finite neighborhood), and in some special cases, leading to a completely local effect.
This significantly simplifies the calculation in the Green's function based discrete dislocation dynamics method, in which a linear system has to be solved over the entire system for the long-range effect of vacancy bulk diffusion, and the long-range Peach-Koehler climb force still has to be calculated.

We have also generalized this continuum formulation to include the pipe diffusion-assisted self-climb based on the discrete self-climb dislocation dynamics model~\cite{Niu2017,Niu2019}.  This continuum dislocation climb velocity can be applied in any available continuum dislocation dynamics frameworks, and we present an implementation of this continuum climb formulation in the continuum dislocation dynamics framework using dislocation density potential functions (DDPFs) \cite{zhuyichao2015continuum,Niu2018}.

The obtained continuum climb velocity formulation applies to dislocations of a single slip system. It can be used in dislocation density based continuum dislocation dynamics models for the averaged climb behavior, i.e. climb of the geometrically necessary dislocations.

\section*{Acknowledgments}
Y Xiang was supported
by the Hong Kong Research Grants Council Collaborative Research Fund C1005-19G and the Project of Hetao Shenzhen-HKUST Innovation Cooperation Zone HZQB-KCZYB-2020083.
YJ Gu was supported by the Structural Metal Alloys Program (A18B1b0061) of A*STAR in Singapore.
SY Dai was supported by National Natural Science Foundation of China Grant No.12071363.
XH Niu's research is supported by National Natural Science Foundation of China under the grant number 11801214 and the Natural Science Foundation of Fujian Province of China under the grant number 2021J011193.

\bibliography{ref}
\bibliographystyle{plain}

\end{document}